\documentclass[acmsmall]{acmart}
\usepackage{booktabs}
\usepackage{graphicx}
\usepackage{colortbl} 
\usepackage{footnote}
\usepackage{makecell}
\usepackage{multicol}

\AtBeginDocument{%
  \providecommand\BibTeX{{%
    \normalfont B\kern-0.5em{\scshape i\kern-0.25em b}\kern-0.8em\TeX}}}

\setcopyright{cc}
\setcctype{by}
\acmJournal{PACMHCI}
\acmYear{2025} \acmVolume{9} \acmNumber{7}
\acmArticle{CSCW492} \acmMonth{11}
\acmDOI{10.1145/3757673}

\begin{document}

\title{The Human Labour of Data Work: Capturing Cultural Diversity through \textsc{World Wide Dishes}}

\author{Siobhan Mackenzie Hall}
\authornote{Joint first authors.}
\affiliation{%
  \institution{University of Oxford and Oxford Artificial Intelligence Society}
  \country{United Kingdom}
  }
\email{siobhan.hall@nds.ox.ac.uk}

\author{Samantha Dalal}
\authornotemark[1]
\affiliation{%
  \institution{Department of Information Science, University of Colorado Boulder}
  \country{USA}
  }
\email{samantha.dalal@colorado.edu}

\author{Raesetje Sefala}
\affiliation{%
  \institution{School of Computer Science, McGill University and Distributed Artifical Intelligence Research Institute}
  \country{Canada}
  }
\email{raesetje.sefala@mail.mcgill.ca}

\author{Foutse Yuehgoh}
\affiliation{%
  \institution{KmerAI}
  \country{France}
  }
\email{foutse@kmer-ai.org}

\author{Aisha Alaagib}
\affiliation{%
  \institution{Independent researcher}
  \country{Saudi Arabia}
  }
\email{aalaagib@aimsammi.org}

\author{Imane Hamzaoui}
\affiliation{%
  \institution{École nationale Supérieure d'Informatique Algiers and New York University Abu Dhabi}
  \country{Algeria and United Arab Emirates}
  }
\email{ji\_hamzaoui@esi.dz}

\author{Shu Ishida}
\affiliation{%
  \institution{Autodesk and Oxford Artificial Intelligence Society}
  \country{United Kingdom}
  }
\email{shu.ishida@oxon.org}

\author{Jabez Magomere}
\affiliation{%
  \institution{University of Oxford and Oxford Artificial Intelligence Society}
  \country{United Kingdom}
  }
\email{jabez.magomere@keble.ox.ac.uk}

\author{Lauren Crais}
\affiliation{%
  \institution{Faculty of Law, University of Oxford and Oxford Artificial Intelligence Society}
  \country{United Kingdom}
  }
\email{lauren.crais@law.ox.ac.uk}

\author{Aya Salama}
\affiliation{%
  \institution{Independent Researcher}
  \country{Egypt}
  }
\email{aya0salama@gmail.com}

\author{Tejumade Afonja}
\affiliation{%
  \institution{CISPA Helmholtz Center for Information Security and AI Saturdays Lagos}
  \country{Germany and Nigeria}
  }
\email{tejumade.afonja@cispa.de}

\renewcommand{\shortauthors}{Siobhan Mackenzie Hall et al.}

\begin{abstract}
This paper provides guidance for building and maintaining infrastructure for participatory AI efforts by sharing reflections on building \textsc{World Wide Dishes (WWD)}, a bottom-up, community-led image and text dataset of culinary dishes and associated cultural customs. We present \textsc{WWD} as an example of participatory dataset creation, where community members both guide the design of the research process and contribute to the crowdsourced dataset. This approach incorporates localised expertise and knowledge to address the limitations of web-scraped Internet datasets acknowledged in the Participatory AI discourse. We show that our approach can result in curated, high-quality data that supports decentralised contributions from communities that do not typically contribute to datasets due to a variety of systemic factors. Our project demonstrates the importance of \textit{participatory mediators} in supporting community engagement by identifying the kinds of labour they performed to make \textsc{WWD} possible. We surface three dimensions of labour performed by participatory mediators that are crucial for participatory dataset construction: building trust with community members, making participation accessible, and contextualising community values to support meaningful data collection. Drawing on our findings, we put forth five lessons for building infrastructure to support future participatory AI efforts.

\end{abstract}



\received{October 2024}
\received[revised]{April 2025}
\received[accepted]{August 2025}

\maketitle

\section{Introduction}\label{introduction}

Large-scale Machine Learning (ML) models underpin many essential digital services we interact with on an everyday basis, such as search engines, movie captions, and image generation tools. These ML models are socio-technical systems that both shape and are shaped by broader societal structures \cite{star1990power,pinch1984social}. As a result, ML models inherit, reproduce, and amplify the values, norms, and biases of the societal structures in which they are developed. These models are imbued with social values through the datasets they are trained on and evaluated against \cite{birhane2021large,schuhmann2022laion}. These datasets are typically constructed by scraping available data from the Internet~\cite{cartografias_internet2025, mwema_birhane2025_internet, goodrobot_mitchell2025_internet}. The Internet, however, is not a neutral repository of human knowledge: it disproportionately represents perspectives from the Global North and reflects existing social, political, and economic power hierarchies \cite{luccioni2024stable_bias,hongevaldatarace23}. ML models built using web-scraped datasets tend to perpetuate harm against communities that are underrepresented in training and evaluation data, creating a cycle that reinforces existing biases \cite{magomere2025worldwiderecipecommunitycentred,katzman2023taxonomising,qadri2025case}.

To address representational gaps in training and evaluation data, ML researchers are actively exploring participatory approaches that directly engage marginalised communities in dataset development\cite{singh2024aya_dataset,kirk2024prism,ramaswamy2023geodegeographicallydiverseevaluation}. These efforts aim to address the ``long-tailed’’\cite{massiceti2021orbit} distribution problem---where minority perspectives are vastly outnumbered relative to the overall distribution of data---by engaging historically marginalised communities in dataset development. Building on recent advances in community-engaged dataset development \cite{singh2024aya_dataset,massiceti2021orbit}, our work illuminates the socio-technical infrastructure necessary to support community-engaged data curation approaches. We draw on CSCW scholarship investigating the labour underpinning large-scale systems and shift focus from the dataset as the final product to the \textit{data work} involved in its creation \cite{sambasivan2021everyone,scheuermanDatasetsHavePolitics2021}.

We shift focus to the data work involved in building a dataset alongside community members by asking: \textit{How can researchers collect cultural data using a bottom-up, community-led approach?} To answer our research question, we use first-person methods \cite{erete2021,homewood2023self,wirfs2021examining,cifor2019inscribing,desjardins2021introduction,howell2021cracks} to engage in a design retrospective \cite{irani2016stories,borowski2023between} of our process of building the \textsc{World Wide Dishes (WWD)} dataset, a text and image dataset of culinary dishes and their associated cultural customs. The \textsc{WWD} dataset emerged from an effort to combat the ``long-tailed'' nature of datasets where cultural artefacts from marginalised groups are disproportionately underrepresented due to myriad technical \cite{hong2024s} and systemic causes \cite{Noble+2018}. \textsc{WWD} collects data about food due to the cultural significance of cuisine: food is deeply intertwined with history, geography, and religious symbolism \cite{unesco2003convention}. Through our design retrospective, we surface how the labour performed by participatory mediators, whom we here call \textit{Community Ambassadors} (CAs) \cite{singh2024aya_dataset}, facilitates bottom-up, community-led data collection efforts. We find that CAs make bottom-up, community-led data collection possible by doing three key things: building trust with community members, making participation accessible, and contextualising community values to support meaningful data collection. We argue that this labour is necessary to make participatory AI efforts successful in action.

To ground our design retrospective of \textsc{WWD}, we first provide historical context that motivated the development of the \textsc{WWD} project in Section \ref{sec:background}. We then describe the \textsc{WWD} dataset, the technical infrastructure built to facilitate data collection, and the stakeholders involved throughout the project in Section \ref{sec:system}. In Section \ref{sec:methods}, we describe how we reflected on the process of developing and deploying \textsc{WWD} and unpack how our positionality impacts both the creation and analysis of our reflections. Through our retrospective analysis, we surface how CAs make bottom-up, community-led approaches to dataset collection possible. In Section \ref{sec:rollout}, we describe three ways in which CAs perform labour to support the development and deployment of \textsc{WWD}. In Section \ref{sec:discussion}, we provide implications for methods and implementation of future participatory AI efforts, and reflect on how tensions that emerged during the data work were manifestations of deeper structural issues within the AI/ML development pipeline. We conclude our paper with a call for future work to invest in and build the necessary infrastructure to support participatory AI efforts in Section \ref{sec:limitations}. 

We build on emergent efforts in the participatory AI space that develop ML models by community members, for community members, such as Masakhane \cite{orife2020masakhaneoriginal} and Cohere's AYA Project\cite{singh2024aya_dataset}. We extend the participatory design literature in CSCW \cite{irani2016stories,howell2021cracks,le2015strangers,harrington2019deconstructing} by characterising how a participatory AI effort unfolds in practice and making visible the labour performed by participatory mediators throughout the research process. Our paper shows how bottom-up, community-driven ML development efforts are contingent on labour performed by CAs. These participatory mediators are an essential infrastructural component that makes participatory design projects viable. We present suggestions for how future participatory AI efforts can account for and support the labour of participatory mediators.

\section{{Literature Review}}
\label{sec:lit-review}
We explore four key areas related to bias and participation in machine learning systems. We begin by examining the literature on how bias manifests in ML datasets and its consequences, followed by an overview of how researchers and practitioners seek to mitigate bias in datasets via crowdsourcing approaches. We contrast top-down crowdsourcing approaches to bottom-up approaches that offer more inclusive alternatives to traditional crowdsourcing. Finally, we draw lessons from established participatory design traditions in social computing to examine both the limitations and possibilities of participatory approaches in AI development.
\subsection{Bias in machine learning datasets}
In this paper, we define bias as a reflection of human subjectivity \cite{Miceli2022studyingup,scheuermanDatasetsHavePolitics2021}. Data are produced by humans and therefore inextricable from human subjectivity \cite{d2023data,haraway2013situated}. When a dataset is created, people make decisions about what deserves to be captured and how to classify what they are capturing \cite{bowker2000sorting}. Depending on who does the classifying and why the classification project is being undertaken, some people may be accounted for while others are dismissed, invisible in the final dataset \cite{durkheim2009primitive}. We say a dataset is biased when one perspective is disproportionately represented to the detriment of others as a result of these subjective decisions. In this paper, we follow calls by \citet{Miceli2022studyingup} and \citet{Barabas2020} to use bias as an analytical lens for examining datasets commonly used in ML development. This approach allows us to trace bias in ML model outputs back to the subjectivities embedded in datasets and, eventually, questions of who was involved in dataset creation \cite{Barabas2020,Miceli2022studyingup}. 

Datasets, carrying the worldviews and beliefs of their creators, can precipitate real-world harms when ML models trained on them reproduce these worldviews in their outputs \cite{Blodgett2021_bias,birhane2021misogyny,weidinger2023sociotechnical,shankar2017allocational,Scheuerman2019gender,hall2024visogender_bias}. For example, \citet{buolamwiniGenderShades} showed that common facial recognition models often failed to ``see'' Black faces, which could result in a wide variety of real-world harms when facial recognition systems are used to verify workers' identities \cite{watkins2023face} or surveil low-income housing residents \cite{macmillan2023cameras}. \citet{buolamwiniGenderShades} traced the roots of the systems' failures back to the training datasets having lacked adequate and appropriate representation of Black people. Studying these representational lapses reveals the social values and power dynamics that shaped dataset construction, resulting in the systematic exclusion of Black people's faces in digital datasets \cite{Barabas2020}.

Machine learning research has extensively studied harmful model behaviours as they relate to gender \cite{berg2022prompt_bias, hall2024visogender_bias, WinogenderRudinger2018_bias,birhane2021misogyny} and race \cite{currie2024genderethnicity, west2024fieldgenderethnicity,buolamwiniGenderShades}. We respond to calls to expand the investigation of model breakdowns to additional axes \cite{weidinger2023sociotechnical} such as cultural representation. In this work, we use the term ``culture'' to mean ``cultural heritage'': both tangible objects (e.g., monuments, sites) and intangible practices (e.g., traditions, rituals) that have a symbolic relationship to cultural identity ~\cite{adilazuarda2024culturesurvey, blake2000defining}.
We build on \citet{Miceli2022studyingup} and \citet{Barabas2020} to study breakdowns in ML model performance in the context of cultural representation through food. We choose food as a lens into culture because food is a salient cultural artefact. Food is a part of the intangible cultural heritage of a group that is shaped by a region's history, geography, and even religious symbolism.\footnote{Intangible cultural heritage is considered a ``mainspring of cultural diversity'' and the social practices and rituals around food are so deeply intertwined with culture that they are recognised as ``intangible cultural heritage'' under UNESCO's Convention for the Safeguarding of the Intangible Cultural Heritage. https://ich.unesco.org/en/convention. This is the primary international legal instrument in the field.}
Not all cultures, and therefore not all foods, are equally likely to be represented in datasets used in ML applications due to how and by whom these datasets are constructed. We trace absent and flattened cultural representation in model performance back to questions about who participates in dataset construction (see Section \ref{sec:background}).

\subsection{Mitigating bias in machine learning through crowdsourcing representative datasets}
Social computing scholars have identified that datasets can encode biases which ML systems then reproduce \cite{gebru2021datasheets,buolamwiniGenderShades,sweeney2013discrimination,scheuermanDatasetsHavePolitics2021,scheuermanProductsPositionalityHow2024,kapaniaHuntSnarkAnnotator2023,sambasivan2021everyone}. In response to this issue, researchers have begun exploring ways to mitigate bias by crowdsourcing diverse values and expanding the range of worldviews represented in datasets~\cite{singh2024aya_dataset,romero2024cvqa}. Crowdsourcing knowledge tends to take one of two forms: a top-down model that typically involves a corporate intermediary (e.g., ScaleAI, Amazon Mechanical Turk) that contracts community members as data workers and directs their work, or a bottom-up model that directly engages community members in the data collection and does not necessarily rely on a corporate intermediary to contract and oversee community members as data workers~\cite{singh2024aya_dataset}. 

Top-down models of crowdsourcing to build more representative datasets include work such as~\cite{bhutani2024seegull, jha2024visage,rojas2022the}, which pursued diverse and representative samples from the Majority World.\footnote{The term ``Majority World'' is a deviation from the more commonly used literary term ``Global South.'' We have chosen to use ``Majority World'' to highlight that the majority of the populations with which we engage come from these regions.} The ORBIT computer vision dataset~\cite{massiceti2021orbit} demonstrates how top-down models can engage in the ethical sourcing of data contributors, as it involves annotators from Enlabeler (Pty) Ltd,\footnote{https://www.enlabeler.com/} a company that empowers its employees by offering technical skill development alongside their data annotation work. While top-down crowdsourcing can be efficient and more easily implemented at scale, it excludes the many perspectives of people who do not work as data annotators~\cite{geiger2020garbage}. Moreover, top-down data crowdsourcing efforts typically dictate to data contributors what an acceptable submission looks like, without room for bottom-up feedback~\cite{posada2022dispotif}. Top-down models do not always consider practical realities; for example, contributors in top-down models can be flagged for fraudulent behaviour if the company overseeing the crowdsourcing effort suspects that multiple people share an account, a common practice in households where there may only be one computer available~\cite{posada2022coloniality,jones2021refugees}. 

\subsection{Bottom-up participatory opportunities for mitigating bias in machine learning}
Whereas related work has commonly centred attention on top-down crowdsourcing, we focus our attention on bottom-up approaches. Our paper contributes to this discourse by extending the scope and opportunities for participatory methods supporting bias mitigation in ML. Compared to top-down crowdsourcing, bottom-up and community-based crowdsourcing broaden the range of people who can contribute and can empower more people not only to share cultural artefacts, knowledge, and expertise, but also to help shape the parameters of their own participation~\cite{denton2021whose,delgado2023participatory,birhanePowerPeopleOpportunities2022}. We characterise bottom-up, community-based crowdsourcing as efforts that allow for any community member to engage (e.g., they do not need to be formally employed as a data crowd worker), make space for community contributors to actively shape parts of the research process (e.g., they may take part in decisions about what kinds of data should be collected), and tend, in large part, to be volunteer-based. 

While top-down methods for crowdsourcing might be easier to scale, recent efforts to build large-scale datasets for ML applications using bottom-up, community-based crowdsourcing techniques have shown great promise. For example, in 2020,~\citet{orife2020masakhaneoriginal} introduced Masakhane, which is an ongoing open-source, continent-wide online research effort for machine translation for African languages. Masakhane has also consistently demonstrated ethical participatory research design for machine translation and natural language processing in Africa~\cite{nekoto2020masakhane, adelani2022masakhaner,ogundepo2023afriqamasakhane}. Cohere's AYA dataset~\cite{singh2024aya_dataset} also presents a framework for large-scale community data collection. Their collaborators spanned 119 countries and were able to create a multilingual text dataset comprising 513 million instances across 114 languages. All their systems have been open-sourced. The Masakhane and AYA Project initiatives demonstrate that bottom-up, community-based crowdsourcing techniques can yield large-scale datasets that are grounded in community expertise and empower everyday community members to participate in dataset construction without having to be formal data workers.

While community-led efforts, such as the aforementioned initiatives, make significant strides in producing datasets with diverse community-based input, both research efforts foreground the dataset as the primary artefact of interest. In our contribution, we aim to build upon the precedent set by bottom-up crowdsourcing efforts by calling attention to the processes and complexities throughout the data collection process by which these types of datasets are created---such as by designing accessible infrastructure to support data collection and working with communities to build trust and rapport. We demonstrate how our effort, \textsc{World Wide Dishes}, fulfilled our characterisation of bottom-up community-based crowdsourcing by building the infrastructure to support any community member to engage with the project and actively shape the parameters of data collection in Section \ref{sec:system_stakeholders}. 

\subsection{The pitfalls and potentials of participation}
Recognising the limitations of top-down designed datasets, ML researchers have begun a \textit{participatory turn in AI}\footnote{We use the phrase \textit{participatory turn in AI} as a nod to the original authors \citet{delgado2023participatory} who define a shift toward giving stakeholders varying degrees of agency in the design of AI systems.} to construct datasets that represent a wider range of values and identities by granting various degrees of agency to stakeholders throughout the design process~\cite{delgado2023participatory,birhanePowerPeopleOpportunities2022,ParticipationScaleTensions,corbett2023power,suresh2024participation}. Participation is thought to enable better alignment of system performance with end-user expectations, as impacted communities would be involved in shaping the values of the ML systems~\cite{delgado2023participatory}. In other words, communities would be empowered as co-designers of the ML systems that impact them. However, the majority of participatory AI efforts fall short of their promises to support meaningful participation among stakeholders in the design process~\cite{ParticipationScaleTensions}. Researchers argue that inadequate attention is paid to shifting power away from systems designers, who are often the primary beneficiaries of improved model performance, and towards participants who could benefit by ensuring models accurately represent them and encode their values~\cite{birhanePowerPeopleOpportunities2022}. While participatory AI is a relatively nascent subfield within the ML community, \textit{participatory design} is a research tradition that has existed for decades in social computing~\cite{asaro2000transforming,bodker2022can}. In our contribution, we draw upon lessons from participatory design in CSCW to demonstrate how these methods can be applied to achieve meaningful power shifts between researchers and communities. 

CSCW has a long and rich history of leveraging community-based participatory research (CBPR) methods to centre socially marginalised groups' epistemologies in the design of technology~\cite{le2015strangers,sum2023translation,bratteteig2016unpacking,harrington2019deconstructing,bannon2018reimagining,pierre2021getting}. CBPR is used as a method to mitigate the power imbalances between researchers and communities by empowering community members to shape the trajectory and design of research projects~\cite{reasonSAGEHandbookAction2008}. For example,~\citet{asaro2000transforming} documents how, in light of impending technological changes in the workplace, workers themselves were invited to participate in the design of those technologies and thus had the power to shape the impacts of these technologies on their wellbeing.~\citet{harrington2019deconstructing} consider the role of research setting on community access, and therefore the accessibility of participation for targeted groups. In CBPR, practitioners strive to overcome their status as outsiders by building trust through affective and moral connections with community members~\cite{le2015strangers,mcmillan1986sense}. Taken together, CBPR literature overwhelmingly stresses the need for researchers to become imbricated with the communities they are studying; through building these relationships, they become invested in the outcomes of the research from an insider, community member perspective. 

While participatory methods have the potential to give communities a greater say over the design of technology, many of these efforts fall short in meaningfully shifting power from the researchers to the researched~\cite{birhane2021misogyny,ojewaleAIAccountabilityInfrastructure2024,ParticipationScaleTensions}. Participation takes work on behalf of communities and the process of participating can be negatively impactful for participants. For example, \citet{epstein2008rise} critiques participatory design efforts for their tendency to ``other'' socially marginalised groups in the course of trying to single them out for inclusion in research.~\citet{pierre2021getting} document how communities assume the ``epistemic burden'' of producing data about their lived experiences in participatory research efforts.~\citet{Puri01042007} expand upon how participatory design efforts in rural communities require careful attention to building support structures that lower the barrier to entry and build intrinsic motivation within community members to engage in participation. Building on \cite{Puri01042007}, we call attention to the role of \textit{participatory mediators} in building pathways for equitable community participation in dataset development. In CSCW literature, mediators are shown to be a crucial component of technical system adoption in organisations \cite{okamuraHelpingCSCWApplications1995,friedman1989computer,grudin1988cscw}. We extend this concept of mediators to a participatory design project, showing how community ambassadors can be used to mediate interactions between community members and the broader research project throughout its development and deployment phase. We argue that participatory mediators are a crucial component of participatory efforts, especially when those efforts are geared towards including people from socially marginalised communities, and propose design recommendations for implementing future participatory AI projects.

\section{Placing \textsc{World Wide Dishes} in Historical Context}\label{sec:background}

As described in Section \ref{sec:lit-review}, there are many concerns with biases in large-scale machine learning models which can be traced to problems in the underlying training and evaluation datasets. In this section, we explore the issue with long-tailed data distributions, where socially marginalised perspectives are disproportionately underrepresented, and their relationship to bias in ML. We discuss how historic and systemic factors have resulted in these long-tailed distributions. As a result, ML models struggle to learn meaningful patterns or produce accurate, fair results for socially marginalised groups because long-tailed data comprise a negligible portion of the training instances \cite{zhang2023deeplongtailed, massiceti2021orbit}. 

The path for data to be included on the ``Internet'' is not equitable. Data from socially marginalised groups are less likely to be preserved in a digitised format and available on the Internet~\cite{Noble+2018}. Additionally, these data are systematically screened out of training datasets by state-of-the-art filtering models, such as CLIP \cite{radford2021learning}, which are biased towards including data from and about Western cultures~\cite{hong2024s}. Researchers use filtering methods such as CLIP to curate datasets to improve model quality by removing noisy, irrelevant, or potentially harmful content, aiming to optimise the relevance and accuracy of the data for the model’s intended use~\cite{fang2023data}. However, these filtering methods often encode and amplify existing biases, leading to datasets that inadvertently exclude non-Western cultures and marginalised groups \cite{qadri2025case}. Ironically, while these filters are designed to protect against inappropriate or harmful content, they sometimes fail to remove problematic material such as foul language, racism, or harmful stereotypes~\cite{birhane2021misogyny}. This inconsistent filtering reinforces disparities in representation and increases concerns about the long-tailed nature of the datasets.

The root causes of the long-tailed nature of data about socially marginalised groups are intimately intertwined with social, political, and economic histories. In other words, the systemic inclusion from---and misrepresentative nature of---datasets is not solely a product of technical failure. Rather, it is a manifestation of societal values contributing to choices made about whose cultures are recorded and whose are erased~\cite{bowker2000sorting,cheney2017we,brubaker2011select,benthall2019racial,benjamin2019race}. For example, the exclusion and flattening of unique cultures from the African continent in digital media can be traced back to colonial regimes that sought to erase and thereby dehumanise the people living on the continent~\cite{faloyin_africa_2022}. During colonial times, power structures meant that the historical record was kept by Europeans colonising the continent~\cite{bowker2000sorting}. In an attempt to justify their actions, they systematically flattened and dehumanised the lived experiences of those they sought to subjugate. Colonial classification regimes flatten important cultural and geographic differences~\cite{das2022,prabhakaran2022cultural,das2021}. Borders were drawn without regard for existing complex social and ethnic groups, and the subsequent oppression and power dynamics involved in recording history lent themselves to a systematic erasure of distinct experiences~\cite{bowker2000sorting}. More recent depictions of Africa in the international media and popular culture have made little attempt to capture its deeply complex and rich landscape~\cite{faloyin_africa_2022}.

Acknowledging the systemic and historical reasons for this type of flattening, \textsc{WWD} explicitly seeks to counteract it through on-the-ground community consultation and collaboration. The granularity required to capture a region's knowledge is most likely best understood through extensive consultation with communities with lived experiences in those regions. For example, borders may not provide the best guide for demarcating cultural boundaries, and a representation of “Kenya” or “Nigeria” may not be granular enough when we consider the distinct cultural groups within these borders. Some borders may also be misleading: for example, the Semliki River has changed course numerous times over the past few decades, meaning that some cultural groups have found themselves flipping between national identities of Uganda and the Democratic Republic of Congo, depending on a naturally evolving geographical marker~\cite{faloyin_africa_2022}. 

We present the African continent as an example to explain the deep-rooted causes of and potential for the type of cultural erasure that exists across the globe. \textsc{WWD} and the infrastructure developed therein seeks to counteract cultural erasure stemming from limitations in web-scraped datasets by encouraging data submissions from around the world in a granular manner, going beyond national representation by explicitly requesting fine-grained regional details about dishes and their associated cultural customs. We focused on food because it is deeply intertwined with a region's history, geography, and cultural practices. Preserving accurate and granular information about food is a crucial element in recording information about culture. For example, a large-scale dataset of dishes maps one singular dish called ``pottage'' onto eight different African countries, when in fact each of these countries is home to a specific, discrete version of the dish with unique ingredients \cite{winata2024worldcuisines}. This kind of generalisation demonstrates how cultural nuance is lost when datasets fail to capture regional distinctions, reinforcing the need for granular approaches such as \textsc{WWD}~\cite{magomere2025worldwiderecipecommunitycentred}. In the coming sections, we expand on the dataset we created to counteract the flattening of cultural representation and the process of creating this dataset alongside community members.

\section{\textsc{The World Wide Dishes System}}\label{sec:system}

World Wide Dishes is an initiative to build a dataset of objects of cultural significance alongside community members from socially marginalised groups to counteract the historic and systemic flattening and exclusion of their representation. In this section, we describe the \textsc{WWD} dataset, the stakeholders involved in the creation of the dataset, and the architecture of the technical system we built to support data collection.

\subsection{The \textsc{World Wide Dishes} dataset}

\begin{table}[h]
\centering
\caption{\small Data collected through the  \textbf{World Wide Dishes} contribution form.}
\label{tab:dish_questions}
\small
\begin{tabular}{ll}
\toprule
\textbf{Question} & \textbf{Data entry} \\
\midrule
Image and caption & Image upload and text caption \\
Dish name in a local language & Short text \\
Name of local language & Short text \\
Country / countries & Dropdown and free text \\
Region & Free text \\
Attribution to a specific cultural, social, or ethnic group & Free text \\
Time of day eaten & Multiple choice \\
Dish classification & Free text or dropdown \\
Components, elements, and/or ingredients & Dropdown and free text \\
Utensils & Free text \\
Accompanying drinks & Free text \\
Association with a special occasion & Multiple choice \\
Recipe & URL \\
Any other comments & Long-form free text \\
\bottomrule
\end{tabular}
\end{table}

The \textsc{WWD} dataset is the product of a bottom-up, community-led crowdsourcing effort to collect text and, when possible, image data about regional dishes. The Core Organisers (see Section \ref{sec:system_stakeholders}) behind \textsc{WWD} launched a concerted effort to encourage data contributions throughout April 2024. At the end of April 2024, the WWD dataset contained 765 dishes with detailed metadata. This included 372 images verified by Community Contributors (301 images with a Creative Commons or royalty-free licence, and 71 user-uploaded images (shared with permission)). These dishes are associated with 106 countries (see Table \ref{tab:continent_tally} for breakdown of regional coverage), with dish names in 131 languages. Community Contributors contributed information about a variety of dishes ranging from main dishes to desserts (see breakdown of dish types in Table \ref{tab:type_meal_tally}). Notably, 101 of these dishes share more than one region of association. Dishes in \textsc{WWD} are accompanied by detailed metadata  (see Table \ref{tab:dish_questions} for metadata information). Many of these metadata fields (e.g., time of day eaten, utensils used) are cultural aspects of dishes that the Core Organisers would not have known to capture, if not for direct input from community members. In the following subsection, we give an overview of the stakeholders involved in the development of \textsc{WWD} and a description of how they interacted with each other and with the overall project.

\subsection{Stakeholders in the \textsc{World Wide Dishes System}}\label{sec:system_stakeholders}
\begin{figure}[b]
    \vspace*{-10pt}
    \centering
    \includegraphics[width=\textwidth, trim=3cm 0.5cm 3cm 0.5cm, clip]{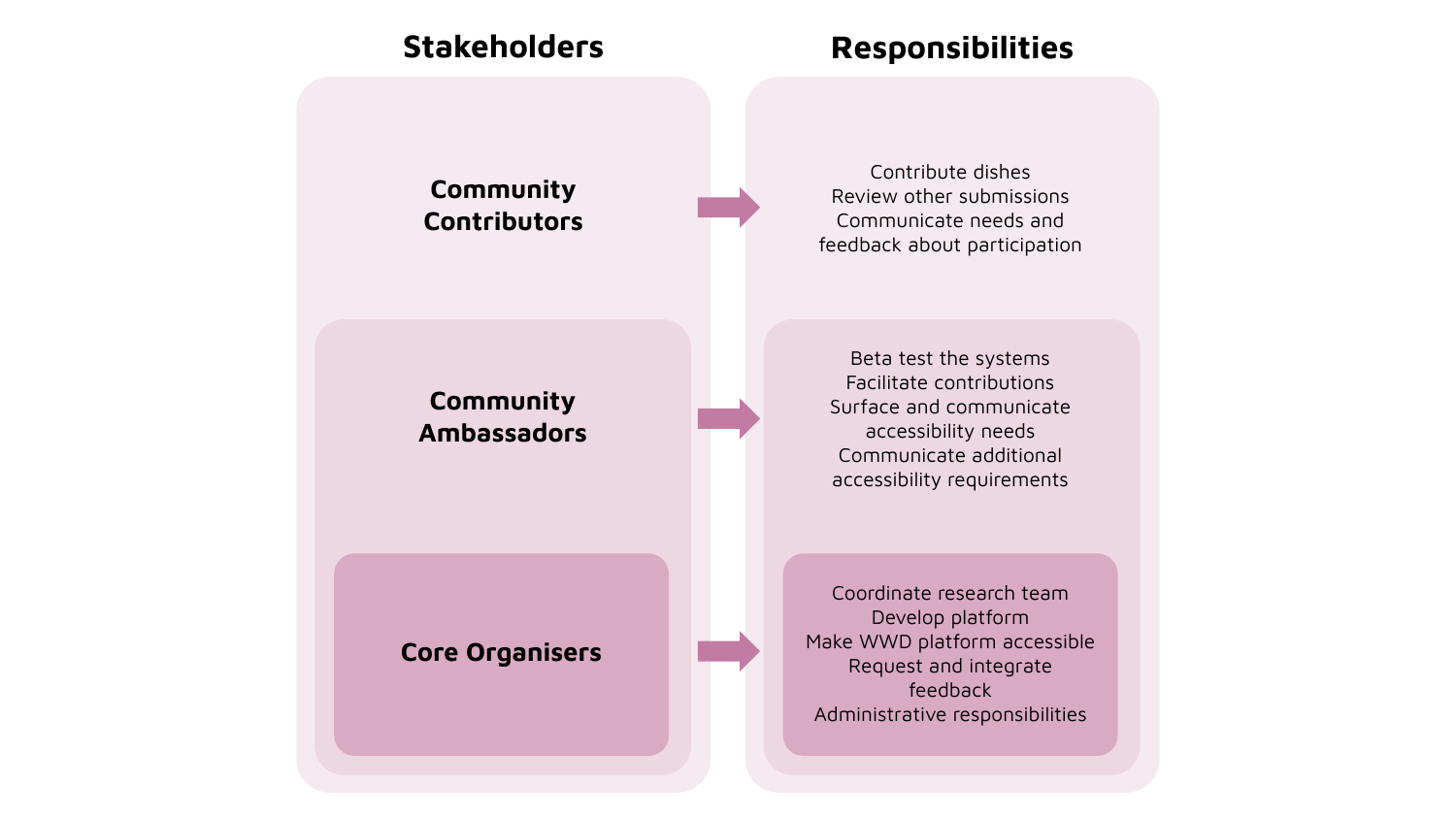}
    \caption{\textbf{Stakeholders in \textsc{World Wide Dishes}}. There are three stakeholder groups in WWD: Community Contributors, Community Ambassadors, and Core Organisers. This figure illustrates the overlapping stakeholder roles in the project.}
    \label{fig:stakeholders}
\end{figure}
\textsc{WWD} is a bottom-up, community-led dataset creation initiative that relied on three key stakeholder groups: Community Contributors (CCs), Community Ambassadors (CAs), and Core Organisers (COs) (Figure \ref{fig:stakeholders}). The stakeholder groups are not mutually exclusive; for example, all COs also served as CAs. CAs were members of the regions that they oversaw (e.g., they were from and maintained close ties with or currently lived in the geographic region). Below we describe the responsibilities of each stakeholder group and how they interacted with each other.

\textbf{Community Contributors (CCs):} CCs accessed the \textsc{WWD} data collection form via the \textsc{WWD} project website and provided cultural knowledge about each dish submitted through the form. CCs also provided feedback on entries provided by other CCs.

\textbf{Community Ambassadors (CAs):} In addition to acting as CCs, CAs distributed the \textsc{WWD} website to and through their social networks to solicit contributions. They hosted focus groups and sometimes filled out the data collection form on behalf of CCs who faced issues with submitting the form themselves, as well as provided translation assistance where needed to help CCs complete submissions in English. CAs also helped to review contributions for dishes from their regions. They further served as a communication channel between CCs and COs, surfacing CC concerns as they arose and communicating subsequent actions by the COs back to the community. Finally, CAs assisted in the data processing stage as described in Section \ref{sec:sys_dataProc}.

\textbf{Core Organisers (COs):} In addition to the responsibilities of CCs and CAs, COs coordinated and handled the administration of the \textsc{WWD} project. They began the initiative to build \textsc{WWD} and developed and iterated the data collection process, front-end \textsc{WWD} project website, back-end database, and data processing pipeline in accordance with community feedback. COs ensured accessibility was a top priority and continuously integrated feedback from the community to achieve this goal. Additionally, they helped to review community contributions for dishes from their regions. All COs also served as CCs and CAs for their home regions.

Together, the three stakeholder groups made collecting fine-grained regional dish data possible. The CAs played a crucial role in mediating interactions between CCs and the overall research project. Because CAs were members of the communities for which they served as ambassadors, they had intimate knowledge of the potential barriers to participation that CCs might face. For example, when the data collection instrument for \textsc{WWD} was being developed, CAs raised the need to build a web interface that was compatible with the 2G cellular networks that were the norm for Internet connectivity in their home regions. In the following section, we describe the technical system architecture through which data contributions were made and conclude with an overview of how community members interacted with and shaped the design of this system.

\subsection{{\textsc{World Wide Dishes System}} technical system architecture}
\begin{figure}[t]
    \centering
    \includegraphics[width=\textwidth]{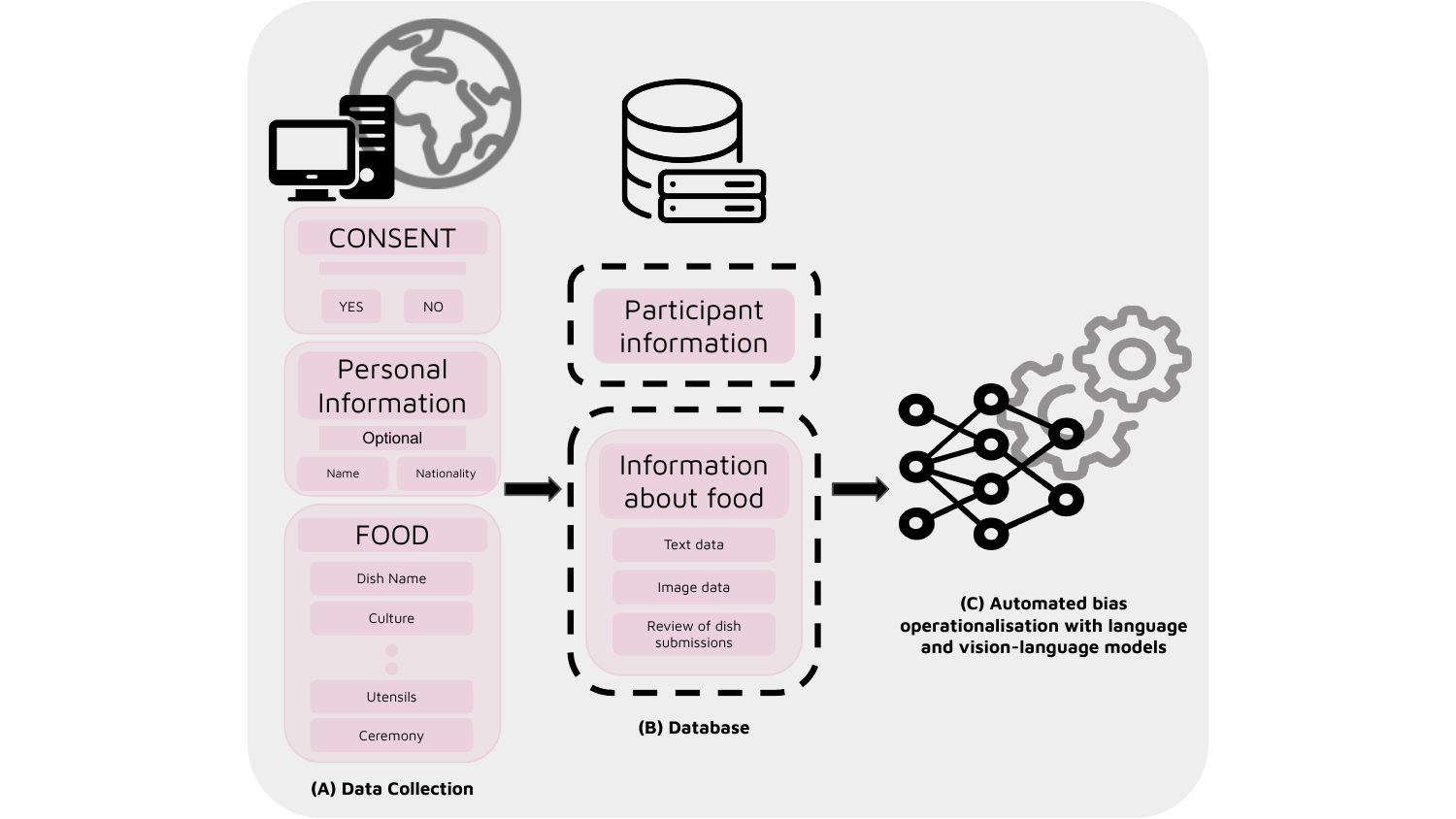}
    \caption{\textbf{Overview of the \textsc{World Wide Dishes} flow}. (A) A Community Contributor accesses \textsc{WWD} through a web browser. They consent to be a research participant and decide whether to create an account or proceed as a guest. They then fill out the data collection form with information about themselves and the dish they submit.  (B) The submission is then stored in the \textsc{WWD} database. We store Community Contributors' information separately from the dish information to preserve their privacy. (C) The full database containing dish information is then used to operationalise bias investigations into generated text and image content using other vision-language and language models.  \textit{All icons in this figure were downloaded from Flaticon. For proper attribution please see Appendix \ref{asec:informed_consent}.}} 
    \label{fig:WWD_Architecture}
\end{figure}

The \textsc{WWD System} is a web application (see Figure \ref{fig:front_page}) built with Django\footnote{https://www.djangoproject.com/} and hosted on Google Cloud Platform, with a PostgreSQL database for storing submissions and Google Cloud Storage for storing image files. Figure \ref{fig:WWD_Architecture} presents an overview of the overall \textsc{WWD System}, including the specific system components. The processes that handle data collection, storage, and processing are described below.

\subsubsection{Data collection}

Data collection occurred through a web-based form accessible on our website (see Figure \ref{fig:front_page}). On the web-based form, CCs were first provided with information about the project and a consent form. If they provided consent and indicated that they were over 18, they were then taken to the data collection form and asked to provide information about their chosen dish. As part of this form, CCs were asked to upload a non-generated image of food from personal records, indicate the dish name in its local language with a translation or phonetic equivalent to English when possible, and provide information about the customs associated with the dish. Please see Table \ref{tab:dish_questions} above for a full list of the fields CCs were asked to provide.

For the \textsc{WWD} data collection process to be useful, contributors had to be recruited and actively engaged. To that end, the COs and CAs worked together to ensure the data collection system was easily shareable via digital social networks to support multiple browsers, and also had interactive elements to amplify CC excitement and motivation.
Recruitment involved the use of many social networks, including through existing communities such as the AYA Project,\footnote{https://aya.for.ai/} Masakhane,\footnote{https://www.masakhane.io/} the Deep Learning IndabaX network,\footnote{https://deeplearningindaba.com/2024/indabax/} AI Saturdays Lagos,\footnote{https://aisaturdayslagos.github.io/} and OLS.\footnote{https://we-are-ols.org/} Posts were also placed on a mailing list to encourage engagement.

\begin{figure}[t]
    \centering
    \includegraphics[width=0.5\textwidth]{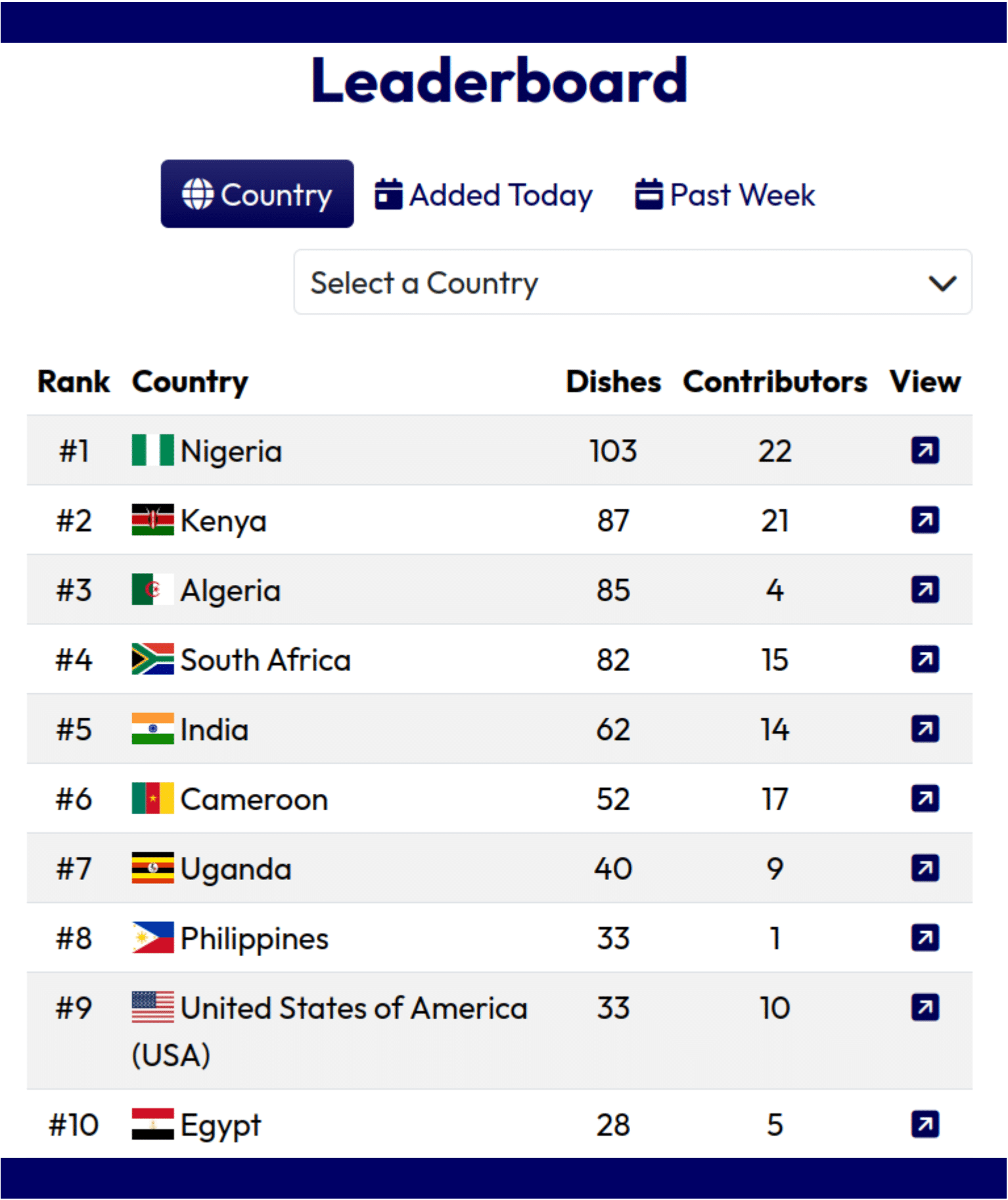}
    \caption{\textbf{Screenshot of the World Wide Dishes Leaderboard showing the top 10 countries by number of dishes contributed, the total number of dishes contributed, and the total number of contributors per country.}}
    \label{fig:leaderboard}
\end{figure}

Engagement was explicitly encouraged through the promotion of a submission leaderboard (see~Figure \ref{fig:leaderboard}),
which displayed the number of dish submissions and contributors received from each country or region. This was done in an attempt to gamify the experience and build excitement and fun. The leaderboard was promoted during recruitment outreach. 

\subsubsection{Data storage}  The collected data is stored in two sections: the dish data, which has been made public and is completely anonymised, and the personal data, which has been collected and stored according to the terms agreed by our ethics review board. We collected names, ages, and national identities, but also accepted anonymous submissions. Names were only used when explicit consent was given in order to acknowledge contributions publicly. Age and national identity data were required to help us understand the demographics of people submitting to \textsc{WWD}. Approval for data collection and the subsequent research study was obtained from the Departmental Research Ethics Committees of the Computer Science Department at Oxford University (reference: CS\_C1A\_24\_004). Notably, the ethics review required us to collect age data, as CCs had to be over the age of 18 to participate; age data was therefore collected even if the CC otherwise submitted dish data anonymously.

\subsubsection{Data processing}\label{sec:sys_dataProc}

\begin{enumerate}
    \item \textbf{Data cleaning} was led by COs to remove duplicate entries within countries and to standardise data entry (i.e. uniform descriptions of regions in a country and language). COs consulted CAs when needed.
    \item \textbf{Translations} were required in special cases. For submissions made by French-speaking CCs from the Democratic Republic of Congo, and to make concessions for accessibility, the COs accepted these specific entries and translated them with an open source machine translation system;  a CO who is a native French speaker then audited these results. Entries were primarily in English, except for some dish names and ingredients without known English translations.
    \item \textbf{Removal of images with uncertain licences} by COs involved the removal of any uploaded images whose licence could not be verified. Accepted images either came from royalty-free sites that did not prohibit their use in ML, had an accompanying Creative Commons licence, or had been personally taken and submitted by a CC with explicit permission given for it to be used for research purposes. 
    \item \textbf{To augment the image data}, COs solicited additional royalty-free and/or Creative Commons images of the submitted dishes from the Internet and consulted CAs for their assessments of the images' accuracy in depicting the submitted dishes.
    \item \textbf{Inconsistencies in submitted data} were handled by consulting CAs where possible, and CAs also consulted with CCs as needed. However, we aimed to collect lived experiences around the world and so we did not make any efforts to police a `ground truth' for each dish, or assign an `origin' or `authenticity' to any dish. Multiple nations or cultural groups can share the same dishes, for a wide variety of reasons including historical trade routes, war and occupation, the redrafting of political borders without regard to cultural borders, and migration patterns. Our task was not to make judgments about cultural `ownership' of the submitted dishes. Instead, the data collection asked participants to offer dishes from their own lived experiences and personal backgrounds, so we expected to see similarities across borders, as well as some reasonable regional and preference variations of the ingredients of dishes within and across borders.
\end{enumerate}

\subsubsection{How stakeholders interacted with and shaped the design of the \textsc{WWD} technical system architecture}
All three stakeholder groups (CCs, CAs, and COs) interacted with and shaped the technical system through which data contributions were made. Stakeholders surfaced additional metadata fields that were important to understanding the traditions of their local dishes during focus groups prior to data collection, and the COs then adjusted the data collection form to include these fields. By guiding the questions asked on the data collection form, community members shaped both the technical data collection system and the resulting \textsc{WWD} dataset. Moreover, COs and CAs who had intimate knowledge of potential barriers to participation for their home regions directly shaped the technical data collection system by advocating for a web interface that was mobile-first and compatible with low-bandwidth cellular networks. CAs provided valuable insight during data collection as to sub-regions and cultures that were underrepresented during data collection and COs and CAs responded by introducing extra efforts to reach these regions.

The \textsc{World Wide Dishes System} is a socio-technical infrastructure that relies both on technical components to collect, store, and process data, and on human actors who animated and shaped the contours of the data collection process. The goal of this system is to develop a fine-grained regional dataset of local cuisines alongside community members by building the social and technical scaffolding needed to support community participation. CCs, CAs, and COs worked together to accomplish this goal. The technical architecture of the \textsc{WWD} System is shaped by the values, norms, and needs of the community members who participated in this effort. In the following section, we describe how members of the Core Organising team reflected on the months-long process of developing and deploying \textsc{WWD} to answer our research question of: \textit{How can researchers collect cultural data using a bottom-up, community-led approach?}

\section{Methods and Analysis}\label{sec:methods}
Our study aims to map out the process of building a dataset of cultural objects (specifically, food) through a bottom-up, community-led approach. In this section, we provide a clear timeline for developing \textsc{WWD}, detail our data collection and reflective practices, and explain our analysis methodology.

\subsection{Timeline and project context}
Planning, building, and deploying the system that resulted in the \textsc{WWD} dataset was a process that occurred over 12 months. 
In October 2023, the first author initiated the \textsc{WWD} project. From November to December 2023, the team expanded to include three more COs, with four more joining in March 2024 (see Table \ref{tab:og_wwd_team} for details about the full team list). The final addition of COs occurred in April 2025 in order to include two CAs in honour of their exceptional commitment to the project. Between January and February 2024, the CO team developed the data collection instrument and website. The WWD website went live in February 2024.
Inspired by the AYA initiative's use of community ambassadors \cite{singh2024aya_dataset}, the COs recruited and onboarded CAs throughout March 2024 to support community engagement. In April 2024, the COs and CAs conducted a month-long campaign for data contributions, resulting in 765 dishes submitted from 106 regions. The research activities conducted as part of the development of \textsc{WWD} were approved by the first author's ethics review panel.

\subsection{Data sources and reflective practice}
Our analysis draws on three key artefacts created throughout the project: 
\begin{enumerate}
    \item \textbf{Documentation artefacts (created November 2023--April 2024):} Throughout the development of the \textsc{WWD} data collection system, all COs, except the second author, contributed to documentation artefacts capturing the development and deployment of \textsc{WWD}. These included a living Google Doc where COs discussed community feedback, the onboarding Google Slide deck for CAs, as well as various documents for planning, collecting, and collating feedback. These documents captured real-time reflections during the development process.
    \item \textbf{Formal paper write-up (May 2024):} Around May 2024, the COs engaged in intentional reflection on the process of developing and deploying \textsc{WWD}, culminating in a paper submitted to a conference that focussed on the \textsc{WWD} dataset and its use for understanding representational harms in foundation models. During the paper writing process, the COs analysed the \textsc{WWD} database (e.g., regional coverage, number of contributions, interesting features of the data) and the documentation artefacts. As the COs continued to discuss the process of developing and deploying \textsc{WWD}, the CAs and the role they played in making this project possible emerged as a key theme. The reflections and documents we produced during the formal paper write-up served as part of our reflective process for this CSCW paper. They helped us to crystallise our focus on CAs and their labour. 
    \item \textbf{Post-mortem reflections (collected August--October 2024):} Four months after data collection concluded, nine COs who also served as CAs produced structured reflections on their experiences. As we show in Figure \ref{fig:stakeholders}, all COs served as CAs. Four COs provided audio recordings through voice notes in private WhatsApp conversations with the second author, who asynchronously asked follow-up questions similar to a semi-structured interview. These recordings were transcribed using MS Word's voice-to-text feature. The remaining five COs wrote detailed reflections in a shared Google Doc. All reflections responded to prompts (see Appendix \ref{asec:prompts} for full list of prompts) created by the second author, who has extensive experience with ethnographic research methods.
\end{enumerate}

Our study is a design retrospective \cite{irani2016stories} reflecting on our process of developing and deploying \textsc{WWD}. To conduct our design retrospective, we draw on established first-person methods in human-computer interaction research \cite{desjardins2021introduction}, which enable researchers to position their reflections on the design process as the key artefact to be analysed. Specifically, we engage in a form of collaborative autoethnography \cite{Mack2021group,chang2021individual} where two or more researchers engage in autoethnographic reflections and dialogue to ``simultaneously generate, interpret, and articulate data about a common phenomenon’’ \cite{cifor2019inscribing}. Next, we provide more detail on our analysis of our artefacts and how we arrived at our findings.

\subsection{Analysis}
We employed a retrospective analysis approach \cite{wirfs2021examining,howell2021cracks} to examine our experiences in designing and deploying \textsc{WWD}. Because the events leading up to the creation of \textsc{WWD} have already passed, the data we analyse are reconstructions of our design process in the form of the artefacts we outline above \cite{wirfs2021examining}. Intentional retrospection of our process of deploying \textsc{WWD} took place between November 2023 and May 2024. This resulted in documentation artefacts and dialogic exchanges between the COs that contributed to a formal paper write-up for an ML venue's dataset track. After the formal paper write-up was completed in May 2024, it became clear to the COs that there was much more to unpack about the social processes that made \textsc{WWD} feasible. Therefore, in July 2024, the COs decided to write a separate research paper that unpacked the methodology underpinning \textsc{WWD} with a particular focus on the role of CAs.

To reflect on and analyse the role of CAs, nine of the eleven COs who joined this paper as authors engaged in post-mortem reflections on their experience serving as ambassadors. The second author, who was brought onto the research team for her expertise in CSCW literature and experience conducting ethnographic research, provided prompts to facilitate the COs’ reflections. Prompts included open-ended questions such as: \textit{``Can you reflect on why you think contributors engaged with you?’’} and \textit{``Can you describe how you introduced the project to your community?''} (see Appendix \ref{asec:prompts} for full list of prompts). Postmortem reflections were generated throughout September and October 2024. The first and second authors then coded reflections from three of the COs, and finally conducted a card-sorting exercise to develop initial themes. The resulting initial themes were shared with all authors for feedback and to agree upon a coding scheme. The remaining reflections were then coded according to the agreed-upon scheme.

Through our analysis, we identified the crucial role of CAs in developing and deploying \textsc{WWD}. We find that they act as participatory mediators who comprise the \textit{human infrastructure} that animates large-scale participatory design projects. In Section \ref{sec:rollout}, we unpack \textbf{three functions of participatory design research that participatory mediators facilitate: building trust, making research accessible, and contextualising community values and culture}.

\begin{table}[h]
\centering
\renewcommand{\arraystretch}{1.5} 

\caption{\small \textbf{The original World Wide Dishes team members and their corresponding positionality.}}
\label{tab:og_wwd_team}
\small 
\begin{tabular}{p{1cm}p{2.25cm}p{2.5cm}p{2cm}p{1.5cm}p{1.5cm}} 
\toprule
\textbf{Code}& \textbf{Background}  &  
\makecell[l]{\textbf{Involvement} \\ \textbf{in the} \\ \textbf{African ML} \\ \textbf{community}} & 
\makecell[l]{\textbf{Research} \\ \textbf{Institution}} &
\makecell[l]{\textbf{Community} \\ \textbf{Ambassador}} &
\makecell[l]{\textbf{Length of} \\ \textbf{involvement}} \\ 
\midrule
O1  & Health Sciences and Machine Learning & Deep Learning Indaba  & University of Oxford & South Africa & 12 months \\
O2  & Law and Cultural Heritage Studies & -- &  University of Oxford &United States & 11 months  \\
O3  & Engineering and Machine Learning & --  & University of Oxford & Japan & 11 months  \\
O4  & Computer Science and Machine Learning & Deep Learning Indaba & University of Oxford & Kenya & 11 months \\ 
O5  & Mechanical Engineering and Computer Science &  AI Saturdays Lagos, Deep Learning Indaba & CISPA Helmholtz Center for Information Security & Nigeria & 8 months \\
O6  & Machine Learning & Deep Learning Indaba, Deep Learning IndabaX Egypt  & Independent Researcher & Egypt & 8 months \\
O7  & Mathematics and Computer Science & Deep Learning Indaba  & Independent Researcher & Sudan & 8 months \\
O8 & Mathematics and Computer Science & Deep Learning Indaba, Deep Learning IndabaX Cameroon, KmerAI  & Conservatoire National des Arts et Métiers & Cameroon & 8 months  \\
O9  & Machine Learning & Deep Learning Indaba, African Computer Vision Summer School  & DAIR, McGill \& MILA  & South Africa & 7 months \\
O10  & Computer Science and Machine Learning & Deep Learning Indaba, Deep Learning IndabaX Algeria  & École nationale Supérieure d’Informatique Algiers & Algeria & 7 months \\ 

\bottomrule
\end{tabular}
\end{table}

\subsection{Positionality}
Our positionality is crucial to understanding how the data (e.g., post-mortems) were generated, analysed, and interpreted to arrive at the three themes we present in Section \ref{sec:rollout}. We take the position that the data we analysed for our paper are inextricable from the subjectivities of the data producers (e.g., COs) \cite{d2023data}. In this section, we provide an overview of the COs’ positionalities and reflect on how these impact both the development and deployment of \textsc{WWD} and the creation and analysis of the post-mortem reflections.
\subsubsection{Who are the authors?}
Ten of the eleven authors of this paper were COs who created \textsc{WWD} and served as CAs during the development and deployment process. The second author, who was not part of the original team, was brought onto the project specifically for their expertise in social computing research methods and ethnographic approaches. The majority of the author list have lived most or all of their lives in the regions for which they served as CAs, only briefly or recently emigrating to pursue secondary or post-secondary graduate degrees. As such, they maintain strong community connections with their home regions, which enabled them to access local social networks to recruit CCs. Nine of the eleven authors for this paper are ML researchers, and the majority are deeply involved in the African ML community, which aims to promote ML development by Africans for Africans. In short, all authors of this paper were, except for the second, members of the communities from which they solicited data contributions and have deep technical expertise in either ML or cultural and culinary heritage, or both. All the authors of this paper, except the second, were active participants in the production of the documentation and formal paper write-up artefacts. 
\subsubsection{How does who we are impact what we did?}
Our positionality significantly shaped how we developed, deployed, reflected on, and made sense of the \textsc{WWD} project. As ML researchers engaging with CCs during the deployment process, we occupied positions of technical authority, which we leveraged by demonstrating how existing technologies inadequately represented local dishes. This likely helped to motivate community participation. Simultaneously, our identities as community members also likely played a significant role in establishing trust and buy-in. As community members, we had intimate knowledge of the relevant cultural norms which helped us to adapt our design of the \textsc{WWD} system to accommodate regional technological constraints and communication preferences.

Our use of first-person methods acknowledges the subjective nature of our experiences as CAs, with our analysis reflecting what we found significant—particularly the emotional and social labour involved in supporting \textsc{WWD}'s development. First-person methods allow us to recognise the produced and subjective nature of our reflections serving as CAs. The choice of topics we focus on in our analysis is coloured by what we felt was important to unpack as community members: the things that weighed heavily on our consciences as we both orchestrated and supported the development and deployment of \textsc{WWD}. We embrace this subjectivity and treat it as an asset that helps us provide a rich, ``thick description'' \cite{geertz2008thick} of the social and emotional labor that participatory mediators perform in supporting participatory design research.

 \section{Building \textsc{World Wide Dishes}: Lessons from the field}\label{sec:rollout}

Building the \textsc{WWD} dataset was a process that took over seven months, and involved twelve COs and more than 170 CAs and CCs. The process of launching the \textsc{WWD} data collection effort involved a beta-testing stage to solicit an initial round of feedback from community members about the data collection instrument and process, as well as hosting ongoing focus groups with CCs to surface issues with knowledge contributions. Through systematic analysis of post-mortems and field notes from the COs, we identify how CAs acted as participatory mediators who made this bottom-up, community-led data collection process possible. In this section, we identify three ways in which CAs made the \textsc{WWD} project possible: building trust between community members and the research process, making participation accessible, and contextualising community values to support effective data collection.

\subsection{Community Ambassadors build trust between community members and the research process}
Collecting long-tailed data requires accessing and connecting with underrepresented populations who tend to be ``hard to reach''.\footnote{We use quotation marks to draw attention to the fact that the term ``hard to reach'' is value-laden. Calling a population ``hard to reach'' redirects attention away from the reasons why that population may be difficult for the researcher to access and instead focuses attention on the group as the problem, rather than on the institution or researcher as the potential problem. For more, see~\cite{benjamin2016informed,epstein2008rise}.} Obtaining this access typically requires working with community gatekeepers who can serve as intermediaries between the researchers and community~\cite{le2015strangers,epstein2008rise}. However, the presence of a community gatekeeper\footnote{We use the term ``gatekeeper'', as it is known in participatory design literature, to refer to those community members who help outsiders interact with community members while preserving the integrity of the community \cite{epstein2008rise, benjamin2016informed}.} in a project does not, in itself, resolve ethical questions around shifting the distribution of power from the researchers to the researched. In building \textsc{WWD}, we made an intentional choice to involve community gatekeepers (e.g., CAs) as members of the research team who shaped the data collection process, the development of the data collection tool, and the paper authorship process. This integration of CAs as core team members rather than mere gatekeepers established a foundation of trust, as community members could see their own representatives helping to shape the research rather than simply facilitating access to them.

The first author, who is a researcher at a prestigious Western university, recognised that their positionality put them in a position of power as relates to the collection of cultural data from communities to which they do not belong. While they are originally from South Africa and have been deeply involved in community efforts within the African ML community, they felt that the \textsc{WWD} project must be co-led by members from communities who were contributing data. Therefore, they invited multiple people in their professional network who were active researchers and volunteers for local AI communities to be co-leads and co-authors on the project. Two of these, O5 and O7, are Nigerian and Sudanese respectively, and were experienced in conducting data collection efforts with communities within their home countries. As O5, O6, and O7 spread information about \textsc{WWD} to their networks, additional community members representing more than eight countries on the African continent signed up to be CAs for the project. CAs leveraged their positionality, relationships, and social capital to make data collection with communities in Africa possible. 

CAs reflected on how their previous work within the African community and being members of the local communities from which they were soliciting data contributions was crucial to the success of \textsc{WWD}. In post-mortem reflections, O5 shared:
\begin{quote}
    ``As a community leader, I've been organising free classes for AI Saturdays Lagos for over six years. Over time, this commitment has built trust within my community, reassuring people that I genuinely prioritise our shared interests. I would like to think that this trust helped community members feel comfortable contributing to the effort, knowing they aren’t being taken advantage of.'' O5
\end{quote}
CAs had established reputations within the communities from which they were soliciting data contributions, enabling them to collect cultural data alongside community members who trusted their intentions. By positioning trusted community members in leadership roles, the project created a bridge of trust between local communities and the academic research process. CCs could see people they knew and respected vouching for and directing the research, making participation feel more meaningful. This dual positioning---as both community gatekeepers and research team members---was crucial for building trust in both directions. CAs could assure their communities that the research process was ethical and beneficial, while also ensuring the COs understood and respected community norms and values.

Established methods of accessing data contributors, such as cold-calling or simply distributing the call to participate on mass broadcast social media channels (e.g., X, formerly known as Twitter, LinkedIn), were arguably less successful, even when it was the CAs themselves who posted the call. Access to, or the ability to access, potential contributors was insufficient. Instead, CAs often had to rely on the relationships they had already established to solicit contributions. For example, O6 reported in their post-mortem that: ``So, I kept sharing and sharing over social media ... but the contributions were not growing as much as I hoped ... So I reach out to people individually.'' As a result, CAs tapped into their network of friends and family to solicit data contributions. However, reaching out to personal networks was not without risk, as CAs were making potentially uncomfortable requests for volunteer work from friends and relatives. 

Data contributions were not financially remunerated. Accordingly, some CAs shared that the lack of financial compensation made them selective in whom they reached out to about the project, as they were essentially asking their friends and family to do volunteer labour. However, despite the lack of remuneration, CAs reported that many of the people to whom they reached out shared a common desire to participate in a project that could make AI systems, such as T2I models, work for them:
\begin{quote}
    ``It's about strengthening African machine learning, it's about empowering Africans, it's about making sure that all of the AI technology works for us all, and ensuring we're part of the builders of it... so maybe through our several grassroots machine learning efforts, we can get to some kind of balance in [power] regarding whose perspectives shape the building of AI technologies.'' O5
\end{quote}

CAs controlled how they, and by proxy the larger \textsc{WWD} team, interacted with their communities. For example, the COs and CAs administered WhatsApp groups within the larger \textsc{WWD} WhatsApp community, where they carried out conversations with CCs in the language of the region they oversaw. CAs had complete oversight over how they wanted to structure conversations within their WhatsApp group and enforce moderation rules that were proposed by the COs. 

CAs had a shared context with the people they asked to contribute data to the project: they spoke the same language as the CCs, had previously established themselves as trustworthy, and shared similar goals for African leadership in ML projects. Through their dual roles, CAs transformed traditional power dynamics in research, creating pathways of trust between communities and the research process. Their embeddedness in both worlds made them effective translators of concerns, values, and goals, ensuring that trust flowed in both directions throughout the project. Building on this, we expand on the need to meet community members where they are and make participation accessible.

\subsection{Community Ambassadors make participation accessible}

Engaging in bottom-up, community-led data collection requires extra efforts on the part of researchers to make participation accessible. In this project, the researchers did so by providing a more accessible informed consent process and hosting ongoing office hours to help CCs learn how to participate and understand the goals and motivations of the \textsc{WWD} project. It is not enough to build a data contribution form that is ``easy'' to fill out; rather, researchers must invest significant amounts of human labour to animate the data collection effort. CAs served as critical accessibility bridges, identifying barriers the researcher team might have missed and advocating for changes that would make participation truly accessible for their community members.

The COs recognised that the CCs may face challenges in terms of Internet connectivity and would likely interact with our website via mobile devices, as also observed in the AYA community data collection initiative \cite{singh2024aya_dataset, avle2018research}. Therefore, the COs devoted significant time to developing the website with specific considerations in mind to ensure the computational burden was not placed on the end-user device. Particular attention was paid to the responsiveness of the website to allow accessibility from different types of devices, and we can report that 62\% of all CCs accessed the website from mobile devices as opposed to desktops. Additionally, in cases where Internet accessibility was not possible at all due to cellular network restrictions, as was the case for our CCs in the Democratic Republic of the Congo, the CA filled in their data contributions on their behalf.

During beta-testing of the website, CAs flagged that while the consent form was accessible---in that it could be translated and read in the local language of contributors---it was not \textit{understandable}. The COs had originally presented the entire consent form, as approved by their institution's ethics review board, as the first step of the data contribution process on the \textsc{WWD} website. Following feedback from CAs, the COs co-designed a concise consent form (see Figure \ref{fig:informed_consent_new}) that laid out key information about the project's purpose, procedures, and data privacy protections. The full consent form was then linked on the consent overview page (see Appendix \ref{asec:informed_consent}). Upon reflection, O1 recounted that consent forms designed and approved in Western academic contexts were inaccessible for the regions in which the CCs were located. The researchers had to engage in translation of the meaning of certain terms, such as ``cookies'', to make informed consent accessible for all participants. This adaptation of consent processes illustrates how CAs functioned as contextual translators, making formal research requirements accessible to community members by reframing them in locally meaningful ways. Without their intervention, this fundamental aspect of ethical research participation would have remained a barrier.

Collecting image data for ML purposes, as \textsc{WWD} does, is a complex process due to concerns about image ownership. Because \textsc{WWD} was intended to be an open-source dataset with Creative Commons licensing, all submitted images either had to have a Creative Commons license (if it wasn't an original) or be an original (e.g., taken by the contributor) photo shared with explicit permission that it could be used for research purposes. These guidelines were presented clearly on the data contribution form; however, CAs noticed early on that CCs were uploading images from the Internet that did not have Creative Commons licenses. As a result, the COs and CAs began hosting regular ``teach-ins'' during ``office hours'' for CCs to attend in order to learn how to properly make a data contribution. O6 explained the role of office hours in making data contribution guidelines accessible to participants:
\begin{quote}
    ``The [office hours] helped ensure our data collection met our requirements and that contributors understood both the licensing requirements and how to properly complete the submission forms, enabling us to collect exactly what we needed for the project.'' O6
\end{quote}
Office hours lowered the barrier to entry for participation by providing a space for CCs to get help filling in the information on their submissions. CAs recalled how CCs would bring ideas about dishes they wanted to submit, but needed help finding an image they could use or verifying other data fields (e.g., utensils used, associated cultural ceremonies) for a dish. In these cases, the office hours attendees would work together to complete a submission collaboratively. This collaborative approach to data contribution represents how CAs created accessibility not just through technical assistance (e.g., navigating the technicalities of image licensing) but by fostering supportive community environments where participation became a shared activity rather than an individual burden. CAs recognised that accessibility isn't merely about interface design but about creating supportive social structures around the participation process.

Office hours also served as a space for CCs to shape the research process. Office hours attendees were encouraged to share feedback about challenges they faced in making data contributions, which the CAs then fed back to the COs. As a result, the COs amended the data collection instrument, produced additional guidelines for participation, and refined their recruitment strategy. O1 recalls how the team wanted to follow so-called ``professional protocol'' and communicate through official and professional media such as mailing lists and social media. The members of the community immediately asked for WhatsApp groups. 
\begin{quote}
    ``As a team, we tried to think through the implications---the most poignant being the lack of a boundary between our own work and professional lives by engaging with our personal WhatsApp applications. However, having already identified how important it was for us to meet the community where they were, and having had similar calls in the past through our work with established communities on the African continent, we made the commitment to host the community groups on WhatsApp, and maintain open communication between ourselves to make sure we could sustainably engage throughout the data collection process.'' O1 
\end{quote}
The switch to WhatsApp groups highlights how CAs advocated for the use of communication channels that were already embedded in CCs' daily lives, rather than requiring them to adapt to unfamiliar platforms. By meeting CCs where they were technologically and culturally, CAs removed significant barriers to engagement that formal academic processes would have inadvertently created. Building on the feedback received from CAs and CCs, the COs quickly realised how much behind-the-scenes effort was going into a single submission. Data are produced through extensive consultation and effort, and the final entry on the website is a distilled form of the rich engagement that led to it. 

Throughout this process, CAs served as essential accessibility facilitators, identifying barriers, advocating for changes, providing direct support, and creating collaborative environments that provided on-ramps for community members to participate. CAs ensured that the research team met CCs where they were, rather than requiring them to develop new expertise in order to participate.

\subsection{Community Ambassadors contextualise community cultural practices, values, and histories to support meaningful data collection}

Throughout the process of building \textsc{WWD}, it became clear that cultural data is produced through social interaction. The concept of data production is not new~\cite{bowker2000sorting,d2024counting}. However, it is often glossed over in ML papers that present novel datasets while only briefly discussing the data work~\cite{scheuermanDatasetsHavePolitics2021} that goes into constructing a dataset. We find that collecting cultural data requires attending to the social processes involved in data production. Through their reflections, CAs identified that cultural knowledge about food exists within social and familial networks.

Contributors actively engaged in conversations with family and friends, accessing familial networks to collect the data needed for a submission. For example, O8 shared one CC's anecdote in their post-mortem where the CC asked their friends and family to help complete data submissions:
\begin{quote}
    ``For the photos and information I couldn’t get myself, I asked certain resourceful people, particularly my parents and friends, to provide the photos and information about the utensils to use and the appropriate time to eat certain meals.'' O8, \textit{summarising what a CC told them}
\end{quote}
While it may be tempting to argue that CCs could simply look up dish information on the Internet to fill out their forms, this was not a viable option for many of our CCs who come from countries where information about their regional cuisines was simply not available online. Familial networks proved to be a more reliable way to get accurate information about local cuisine. O8 recounted an experience where a CC in the community they oversaw encountered difficulty finding information about a regional Cameroonian dish they had grown up eating:
\begin{quote}
    ``For the dishes I struggled to explain, I turned to various sources for help. When possible, I looked online to see how they were made. However, for some dishes, it was difficult to find accurate information. In those cases, I reached out to my mother and grandmother, who were able to share their expertise and guide me through the process.'' O8, \textit{summarising what a CC told them}
\end{quote}
CAs and CCs alike often turned to their familial networks, especially parents and grandparents, to produce the information they needed about a dish through conversations and glimpses into the archives of family recipes. O5 recalled that during their office hours, CCs shared stories about calling their mothers and sisters to get help filling out information for a dish: ``Some [CCs] mentioned that they reached out to their family to learn more about the native dishes they grew up eating because they realise they don't know how to prepare them.''  

Data about food, and how it is connected to a community's culture, is not simply sitting somewhere, waiting to be harvested. Moreover, in the regions in which we were operating, there was little to no reliable information about cultural cuisines available online. As a result, CCs and CAs produced this data through conversations with kin. Rather than treating contributors as isolated sources of information, CAs supported a collaborative knowledge production process that accounted for the distributed nature of cultural knowledge. This approach ensured that the collected data reflected the collective cultural knowledge about food embedded in family relationships and community traditions.

Food is deeply intertwined with cultural practices and, therefore, some of our CAs chose to ground their data collection efforts in major cultural events that coincided with the data collection phase. For example, O6 shared how they deliberately made use of the overlap with Ramadan and Eid al-Fitr to have focussed calls for data contributions. Ramadan and Eid al-Fitr, which are major holidays for Muslims and occurred during our data collection phase, are often celebrated with special dishes holding cultural significance that are only prepared during the holiday period. O6 explained: ``Food plays a great, very important role in Ramadan and so I was very intentional about just taking pictures of whatever food we made for iftar which is the breaking of the fast.'' The food that is prepared during these holidays holds cultural significance and helps tell a part of the story about a community's values and history. CAs understood the cultural contexts in which food exists, strategically aligning data collection with cultural events to capture authentic representations of dishes that hold significant community value. 

Food is complex and dishes often harbour a deeper cultural significance. For example, O6 recalled how during their experience managing data contributions from Egypt, they realised that the classification of ``Egyptian food'' was fuzzy at best:
\begin{quote}
    ``[WWD] made me more conscious about \textbf{what is} Egyptian Cuisine and the influences from other cuisines that have actually, you know, shaped what we eat and what we consider as Egyptian. As it turns out, it's very intersectional.'' O6, emphasis added
\end{quote}
For O6, capturing data about Egyptian cuisine led them to further investigate the origins of their own cultural dishes. In the weeks following the data collection effort, they attended a session about "food as cultural currency" at the RiseUp Summit Egypt 2024 to learn more about Egyptian cuisine through the eyes of local food experts.

O6's reflections on the fuzzy boundaries of cultural cuisines was echoed by various CAs, who consistently recognised that meaningful data collection required moving beyond simplistic national or regional categorisations. The fuzzy boundaries of culinary traditions extended not only to cultural influences but also to the material composition of dishes themselves, revealing how ecological diversity within regions shapes culinary practices in ways that conventional datasets might overlook. O1 recalled how the fuzzy classification of dishes as belonging to a certain region impacted the data collection process:
\begin{quote}
    ``... depending on the region within the country [the dish] was either going to be made from the cassava plant or a kind of mielie meal. It could be the same dish but depending on the region within the same country they would have a different ingredient for the starch [element].'' O1
\end{quote}

Here, O1 identifies that even though a dish may be considered ``the same'' and may be called by the same name across different regions, the actual ingredients in this dish differ and reflect the differing agriculture practices of various regions. These differing agricultural practices are important markers of the local ecosystems and historical practices of farming and cultivation. Therefore, in the data collection process, it became essential to deconstruct dishes into their constituent ingredients, as reflected in the data structure for \textsc{WWD} (see Table \ref{tab:dish_questions} and Appendix \ref{asec:protocol-screenshots}).

In post-mortem reflections, CAs highlighted the importance of recognising the limitations of using national borders to demarcate cultures. The granularity of representation was essential to ensuring that distinct cultures were not lumped together into a single flattened snapshot, as physical borders often do. O5 recalls how she realised that one of the three major ethnic groups in Nigeria---the Hausa community---was underrepresented in the \textsc{WWD} dataset. They reached out to a personal connection whom they knew was a member of that ethnic community who, in turn, helped share the website with their community.

In line with what we posited in the introduction, we reinforce the idea that the value in a dataset is not exclusively conveyed in its final form, but also through the processes of creating it. In this Findings section, we highlight the immense efforts and considerations that went into engaging with our CCs to produce high-quality, granular data about cultural dishes. Processes such as these are slow, iterative, and very hard to scale, but they are necessary to ensure the production of high-quality and diverse datasets that we would like to see reflected in the ML systems that make use of this data.

\section{Discussion: Implications for Designing Infrastructure to Support Participatory AI Efforts}
\label{sec:discussion}
\textsc{WWD} is a socio-technical infrastructure that supports the community-led collection of cultural food data. Community members' needs and experiences actively shaped both WWD's architecture and the types of data collected, ensuring compatibility with existing digital infrastructure and cultural norms. Building the \textsc{WWD System} in a bottom-up, community-led manner required an immense amount of labour. Data did not simply flood in once the system architecture was built. COs and CAs engaged in \textit{data work}: a socio-technical process often overlooked despite its essential role in shaping dataset epistemology and downstream ML performance~\cite{sambasivan2021everyone,ismailEngagingSolidarityData2018,mollerWhoDoesWork2020,scheuermanProductsPositionalityHow2024}. 

Drawing from our experience with WWD, particularly our interactions on the African continent, we surface five lessons for designing infrastructures that support a \textit{Participatory Turn in AI}. WWD is an example of how to build infrastructure for participatory AI that highlights the crucial role CAs play as \textit{participatory mediators}. These mediators performed essential work connecting community members to the research initiative and from their efforts, we derive \textbf{five key lessons for researchers building datasets for ML models using participatory approaches}. These lessons guide how we might move towards more equitable structures of engagement with socially marginalised communities.

\textbf{Lesson Number One: Invest in and support participatory mediators.}
Collecting fine-grained and representative cultural data is exceptionally difficult. Cultures are not bounded by government borders and/or other manufactured, static systems, but rather may extend across regions and be the product of intercultural exchanges~\cite{gupta2008beyondculture}. This, and the number of cultures present in any geographical area, makes knowing what cultural data is underrepresented almost impossible without extensive consultation. As O1 detailed in their reflections, CAs play a crucial role in identifying what communities or regions are underserved. Their fine-grained understanding of community dynamics enables them to recognise who is systematically excluded or marginalised within these contexts. Then, CAs like O5 can tap into their insider identity to build trust and motivate participation, creating pathways for engagement with populations that might otherwise remain inaccessible to outside researchers. CAs therefore act as \textit{participatory mediators} \cite{okamuraHelpingCSCWApplications1995} who play a crucial role not only in managing technical issues but also in understanding the contexts in which these data contribution efforts play out. Calls for equity-driven participatory design projects advocate researchers to consider the roles of historical context, research setting, and commitment to practical outcomes \cite{harrington2019deconstructing,reasonSAGEHandbookAction2008}. We address calls to actualise equity-driven participatory design engagements by developing the role of CAs who act as participatory mediators, contributing contextual knowledge about communities, bringing the research to communities, and using their insider status as community members to advocate for research goals in line with community interests. We call for future researchers to construct formal organisational roles within their projects for participatory mediators, ensuring their labour is both visible and properly supported within research infrastructures \cite{okamuraHelpingCSCWApplications1995,grudin1988cscw,star1999layers}.

\textbf{Lesson Number Two: Researchers should build on existing community-led initiatives.}
Grassroots communities have a long history of engaging in data activism projects where community members utilise datafication as a tool to capture and legitimise their ways of knowing (e.g., epistemologies) \cite{d2024counting,milan2016alternative,collective1980eleven,Harris2024xrds}. ML researchers seeking to mitigate breakdowns in model performance by collecting more data about socially marginalised groups should carefully investigate whether there are existing data activism efforts \textit{within} communities. As \citet{d2024counting} and \citet{pierre2021getting} call for in their work, equity-driven participatory design engagements should centre the community by imagining the scope and purpose of data collection efforts---what do communities want to achieve in collecting data? Building tools for existing community-led initiatives is a step towards centring community ways of knowing in data activism efforts. In our work, we built upon existing research networks that aim to collect and record cultural data about socially marginalised groups, such as the Deep Learning Indaba and AI Saturdays Lagos (see Table \ref{tab:og_wwd_team} for a comprehensive list). Many of the authors of this paper, who served as CAs, are members of these research networks who see the value in (and potential pitfalls of) pursuing cultural preservation through digitisation. We advise that future researchers seeking to engage in participatory AI efforts, especially as they relate to developing datasets and ML systems for the African continent, familiarise themselves with the core values and missions of established local research networks and take steps to understand the space in which they exist and determine if there is room for non-extractive, mutually beneficial engagement. Doing so will enable researchers to better understand current priorities for community-led research and align their efforts to promote more equitable ML models with community definitions of equity and community criteria of success.

\textbf{Lesson Number Three: Accounting for the produced nature of data requires rethinking the top-down paradigm of crowdsourcing data.}
Our CAs shared the myriad ways in which they supported the collective production of cultural data through synthesising input from familial networks and hosting focus groups. We find that cultural data must be produced through cultural practices. Therefore, efforts to crowdsource cultural data should account for and support, rather than penalise, collaboration on a single data entry \cite{posada2022coloniality}. Crowdsourcing platforms should create affordances that enable community members to ``phone a friend'' and reach out to their existing social networks to collectively source knowledge, rather than relying on top-down task allocation rules that determine which workers should be assigned which task \cite{Kittur2013crowd}. Researchers seeking to engage in participatory AI efforts to collect cultural data should be aware of the limitations of top-down crowdsourcing methods that rely on formal data workers and thereby exclude the knowledge of community members who cannot or do not work as data workers. Following reflections from O6, we call for researchers collecting cultural data to be attuned to the relationship between cultural practices and the data they seek to collect. For example, researchers seeking to collect data about material culture should acknowledge how certain religious traditions impact the kinds of material culture that are visible or otherwise available to outsiders for research purposes.

\textbf{Lesson Number Four: Creating open source datasets using bottom-up, community-led crowdsourcing will require downstream labour to ensure compliance with open-source standards.}
Our results revealed tensions around data collection, particularly with regard to image provenance, information accuracy, and participation benefits—highlighting structural problems within the AI/ML pipeline. Recent participatory ML research efforts protect community contributors through restrictive dataset licenses that assign ownership and usage terms~\cite{birhanePowerPeopleOpportunities2022,longpre2024largelicence}. However, for these licenses to be effective, data must have clear provenance. \textsc{WWD} was designed as an open-source dataset with a Creative Commons license for model evaluation, requiring the dataset creators to demonstrate legitimate rights to all included images. Many contributor-submitted images lacked proper licensing, necessitating extensive consultation between CAs and CCs to explain provenance requirements. Submissions with unclear origins had to be deleted, unfortunately erasing valuable cultural knowledge from our dataset. This verification process proved time-intensive and difficult to scale. Enforcing image upload guidelines in a volunteer-based project resulted in a smaller, less representative dataset than desired, though the rigorous provenance verification enabled us to release our dataset as an open-source project with integrity. The open-source standards and Creative Commons licensing, while valuable, place significant burdens on small, community-led projects like \textsc{WWD}. We call for future researchers seeking to create open-source datasets through participatory methods to build infrastructures that support and fund the labour needed to verify that data for their projects can be used.

\textbf{Lesson Number Five: Examine the motivating factors that are presented to participants, and the promises made alongside them.}
Whilst better representation in datasets will improve quality of service to an extent, it is unlikely to undo the centuries of discrimination and identity prejudice that also influence these AI pipelines. Honest engagement can empower contributors to make informed decisions about how they choose to contribute their knowledge \cite{dalal2024provocation}. We intentionally chose not to make use of professional data centre workers,\footnote{Professional data centre workers are those people employed in a centralised manner to perform data collection tasks. Their livelihood is, therefore, connected to the requirement to engage in data contribution, which does not align with WWD goals. Additionally, even had we wished to use data centre workers, we lacked the resources to which a large technology company might have access, such as the ability to engage an ethical outsourcing company (e.g., Enlabler) to recruit and pay data workers.} but instead used a data collection method that would enable us to reach participants other than those employed in a data worker centre, such as older generations and those across a wide socioeconomic range. Although we would have preferred to compensate each CC, we were unable to do so due to a lack of funding. We instead put out a call for volunteers, clearly stating the amount of work to be done, the motivation behind the project, and what we intended to do with the data after collection. CAs also demonstrated differences in the quality of service provided by widely available AI systems, such as underrepresentation of images and information returned by search queries, as means to encourage submissions. As a result, a number of individuals voluntarily signed up to contribute their time and effort to the project. However, despite recruitment successes, CAs also shared conflicting feelings about potentially exploitative and extractive processes when tapping into the intrinsic motivations that may have motivated CCs' involvement. We surface this lesson because the majority of data contributions came from the African continent, and we asked \textit{why this might be}.

The authors speculated about possible reasons for this result, including whether this was a display of ``Ubuntu in action''. The philosophy of ``Ubuntu'' is rooted in the concept of ``familyhood and unity'' and is central to many African cultures. Ubuntu is recognised by various terms across the continent, including \textit{djema’a} (Arabic), \textit{ubuntu} (Zulu), \textit{ujamaa} (KiSwahili), \textit{munhu} (Chichewa), and \textit{unhu} (Shona). However, the reasons for this participation may be more grounded in the reality that there is a growing awareness of the disproportionate cultural exclusion some regions face in the wake of increasingly large-scale AI model development. There is a belief that the historical discrimination and identity prejudice ingrained in Africa, stemming from the colonial experience, may have contributed to a sense of urgency in contributing data (see Section \ref{sec:background}, where we explicitly address the role of colonisation in creating long-tailed data, which amplifies bias in ML models). As a research team, we also recognise that participation in dataset construction is not a guaranteed way to achieve representational justice at the levels of media systems and society \cite{gillespie2014relevance, gillespie2024generative}.

Relying on these models to ``fix'' centuries of intentional cultural erasure and simply presenting the case as ``participate or be excluded'' obfuscates the deeper systemic issues that dictate whose culture gets recorded and represented and overlooks the deeper systemic issues that will likely constrain the efficacy of these technocentric solutions \cite{dalal2024provocation}. Contributors should be empowered to understand the disparate societal factors at play that influence model development as well as their decision to contribute knowledge. We advocate the researchers to engage reflectively and mindfully with the promises they make to encourage these contributions to ensure informed consent is truly given. Engaging with motivations rooted in deep community values (``familyhood and unity''), alongside those arising from systemic injustices, presents an ethical dilemma that we as researchers continue to navigate.

\section{Limitations and Future Work}\label{sec:limitations}

\subsection{Limitations}
The distributed network we relied on provides a challenge for financial payment. We believe that meaningful compensation is an important standard in data collection that relies on community-based knowledge. In preparing this work, we have---and continue to, extensively---engaged with other practitioners and community members to understand best practices. At this moment, we rely on a process of transparency and informed consent rooted in direct attempts to empower participants, such that the practice is less extractive. We maintain a constant communication channel with all CCs (who are, with permission, rightfully acknowledged in Appendix \ref{asec:contributor_acknowledgement}). 

\subsection{Future work}

Finally, we conclude with suggestions for how future work can engage in infrastructure to support data work. 

The work presented here is not meant to be one size fits all, but rather acts as a starting point from which to change the way we approach ML datasets and position their importance in terms of the \textit{human processes} behind them. Data collection that relies on community input results in high-quality data, but the process involves a lot of labour, and is slow, iterative, and very hard to scale by any other means than what we propose here. We hope that other researchers use this as a foundation for future work.

As mentioned above, simply creating an access point for data submission does not mean that data will easily flow in. Complete infrastructure in the future should involve both accessible technical infrastructure as well as human infrastructure to support direct engagement with stakeholder communities. Additional efforts should support increased accessibility that supports submission in a Contributor's local language to increase accessibility.

Data has value, and promises of ``increased representation'' should be critically assessed to make sure the promises made to CCs are sincere. Improvements can come from increasing infrastructure for payments in a decentralised and distributed network that allows for \textit{meaningful} compensation for data labour, and we continue to advocate for CCs to be the owners of their data, as well as for maintaining~\cite{tonja2024inkubalm} transparency in the process.

\section{Conclusion}\label{sec:conclusion}

Through \textsc{World Wide Dishes}, we present guidance for building and maintaining infrastructure for participatory dataset curation that relies on a bottom-up, community-led effort that can lead to high-quality data enriched with local, community expertise. This enrichment comes from extensive human efforts that rely on engagement with social networks working together to produce cultural data. This was made possible by engaging with existing community networks supported by participatory mediators we refer to as \textit{Community Ambassadors}. These networks are built through trust and careful communication. The infrastructure \textsc{WWD} developed increased accessibility through informed consent processes, careful consideration of accessible language, and the possibility of submitting in multiple ways: directly from CCs or with the assistance of CAs. CAs played a pivotal role in supporting the process by distributing the website, answering questions, and providing translations where needed. While this process is not without its limitations, we expand upon the tensions and present five lessons learned in Section \ref{sec:discussion}, with specific suggestions for future work in Section \ref{sec:limitations}. Data are produced, not found; future efforts to support bottom-up, community-led data collection must therefore provide infrastructure that supports the social and relational processes of data production.

\section*{Acknowledgements}
The authors would like to thank the following people for their feedback and insight during the development of \textsc{World Wide Dishes}: Kavengi Kitonga, Nari Johnson, Rida Qadri, Michael Leventhal, Rishi Vanukuru, Krishna Akhil Kumar Adavi, Julia Dean, Emma Ruttkamp-Bloem, Jordan Wirfs Brock  and Morgan Scheuerman. We also acknowledge the significant contribution of those who contributed local expertise to the development of \textsc{World Wide Dishes} and the analyses we present here. We acknowledge these contributions in~\ref{asec:contributor_acknowledgement}. This work has been supported by the Oxford Artificial Intelligence student society. SI was funded by the Ezoe Memorial Recruit Foundation until March 2024. TA is partially supported by ELSA - European Lighthouse on Secure and Safe AI funded by the European Union under grant agreement No. 101070617. For computing resources, the authors are grateful for support from the Oxford Internet Institute, the OpenAI API Researcher Access Programme, as well as for the generous support from Jonathan Caton and the Google Cloud and Google's Compute for Underrepresented Researcher's Programme.

\bibliographystyle{ACM-Reference-Format}
\bibliography{references}
\clearpage

\appendix
\begin{center}
	\textbf{\Large Appendix}
\end{center}
\addtocontents{toc}
\protect\setcounter{tocdepth}{0}
\tableofcontents

\clearpage
\addtocontents{toc}{\protect\setcounter{tocdepth}{1}}

\renewcommand\thefigure{\thesection.\arabic{figure}}  
\setcounter{figure}{0}
\setcounter{table}{0}
\renewcommand{\thetable}{\thesection.\arabic{table}}

\section{Contributors to the \textsc{World Wide Dishes} dataset}\label{asec:contributor_acknowledgement}

The World Wide Dishes dataset is the product of committed contributors and community ambassadors who were
open and willing to share experience and knowledge close to their homes and cultures, and our dataset would not exist without them. This data belongs to those who added to it, and we believe this recognition is paramount. Not all of our
participants agreed to be recognised in print; all names below have been included with express permission.

\subsection{Distinguished contributors} \label{dist_contributors}

\setlength{\columnsep}{0.5cm}
\begin{multicols}{3}
[
This section is dedicated to our Community Ambassadors, who demonstrated significant and extraordinary commitment to the project, involving time, energy, and—most importantly—cultural expertise.
]

\textit{Albert Njoroge Kahira}

\textit{Ashne Billings}

\textit{Awa-Abuon Fidelis}

\textit{Benazir Kemunto}

\textit{Borel Sonna}

\textit{Bouthina Ikram Zergaouina}

\textit{Carol Topping}

\textit{Cynthia Amol}

\textit{Danilo Jr Dela Cruz}

\textit{Dineo Thobejane}

\textit{Djenki Amina}

\textit{Eliette Mbida}

\textit{Elodie Ngantchou Kemadjou}

\textit{Fridah Mukami Miriti}

\textit{Godwill Ilunga}

\textit{Guiwuo Olive}

\textit{Hennane Douaaelikhlas}

\textit{Ines Bachiri}

\textit{John Wafula Kituyi}

\textit{Kavengi Kitonga}

\textit{Keriann V Engle}

\textit{Kholofelo Sefala}

\textit{Lesego Seitshiro}

\textit{Lynda Ouma}

\textit{Meriem Hamzaoui}

\textit{Mureille Laure Obaya}

\textit{Ngatcheu Nguemeni Pascaline}

\textit{Nicholas Ginsburg}

\textit{Rachel M Hurwitz}

\textit{Robert Bork III}

\textit{S E Harburg-Petrich}

\textit{Samuel Ekuma}

\textit{Sandra Marion Kam Tsemo}

\textit{Temitope Fabiyi}

\textit{Trey Topping}

\textit{Volviane Saphir Mfogo}

\textit{Yousra Ferhani}

\end{multicols}
\setlength{\columnsep}{0.5cm}
\begin{multicols}{3}
[
\subsection{Special mentions}
This section is devoted to our Contributors who shared significant amounts of information about the dishes associated with their homes and cultures, collectively making up more than 20\% of the version of the dataset as released in June 2024.
]

\textit{Aswathi Surendran}

\textit{Bruno Ssekiwere}

\textit{Fernanda Gonçalves Abrantes}

\textit{Mahmoud Hamdy Mahmoud}

\textit{Nneoma Jilaga}

\textit{Ondari Laurah Nyasita}

\end{multicols}
\setlength{\columnsep}{0.5cm}
\begin{multicols}{3}
[
\subsection{Contributors}
We are so grateful to all the Contributors who shared their local knowledge with us, making World Wide Dishes possible on such a scale.
]

\textit{Adwoa Bempomaa}

\textit{Albano dos Santos} 

\textit{Aleksandar Petrov} 

\textit{Alena Bubniak} 

\textit{Allan Bahati} 

\textit{Anjali Rawat}

\textit{Annik Yalnizyan-Carson}

\textit{Arinaitwe Rebecca}

\textit{Ayodele Awokoya}

\textit{Bala Mairiga Abduljalil} 

\textit{Caroline Watson} 

\textit{Clémence Bamouni} 

\textit{Cyril Akafia}

\textit{Dolapo Subair}

\textit{Émilie Eliette-Caroline sNGO Tjomb Assembe}

\textit{Faisal Mustapha Muhammad} 

\textit{Fumiko Kano}

\textit{Harry Mayne}

\textit{Haruna Kaji}

\textit{Hazel Chamboko}

\textit{Ian Hsu} 

\textit{Ian Kanyi} 

\textit{Itangishaka John Esterique} 

\textit{Jason Quist}

\textit{Jean Marie John}

\textit{Karen Kandie} 

\textit{Kathryn Hall}

\textit{Kaweesi Patrick} 

\textit{Kevin Otiato} 

\textit{Lufuluvhi Mudimeli} 

\textit{Mbali Mteshane} 

\textit{Mikhail Sondor}

\textit{Mohar Majumdar} 

\textit{Momo Kell} 

\textit{Monsurat Onabajo}

\textit{Montserrat Vallet} 

\textit{Moyahabo Rabothata}

\textit{Mwenyi Enock Mabisi} 

\textit{Nischal Lal Shrestha} 

\textit{Nomsa Thabethe}

\textit{Odeajo Israel}

\textit{Olumide Buari}

\textit{Opeyemi}

\textit{Ouedraogo Gamal}

\textit{Oyewale Oyediran} 

\textit{Paola Fajardo}

\textit{Phillip Ssempeebwa}

\textit{Pratik Pranav}

\textit{Rostand Tchatat} 

\textit{Sabrina Amrouche}

\textit{Salha Elhadi}

\textit{Samuel Ephraim}

\textit{Samuel Oyedun} 

\textit{Sandra Mon}

\textit{Sarah Akinkunmi}

\textit{Sicelukwanda Zwane} 

\textit{Temitope Fabiyi}

\textit{Tanja Gaustad}  

\textit{Tchinda Tatissong Raphaël} 

\textit{Uriel Nguefack Yefou} 

\textit{Wawira Ndwiga} 

\textit{Ximei Liu}

\textit{Yomna Ahmed Bakry}

\textit{Ziliro Jere} 

\end{multicols}

\clearpage
\section{The front page of the website}
\label{asec:website_front_page}
\begin{figure}[h]
    \centering
    \includegraphics[width=0.8\textwidth]{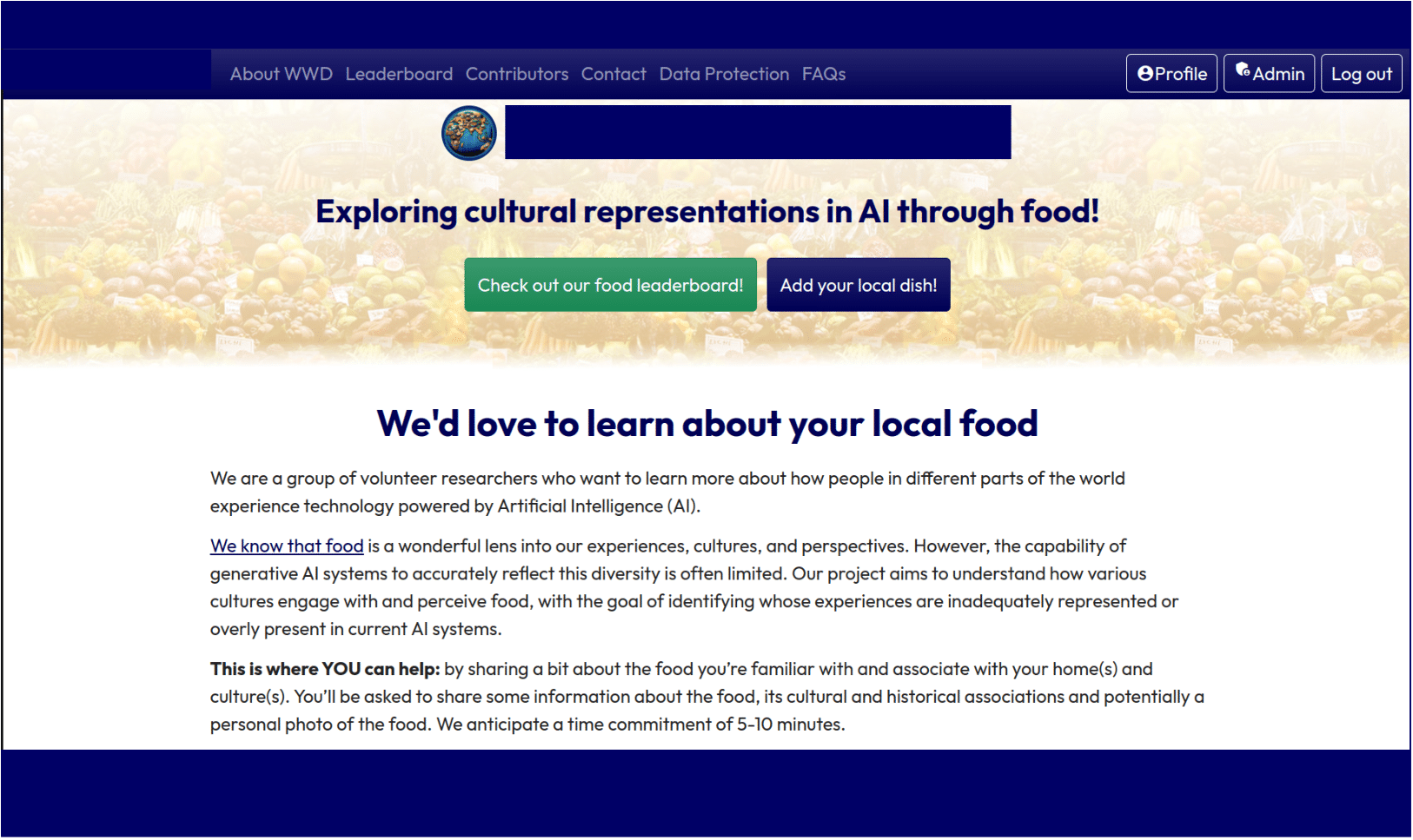}
    \caption{\textbf{Front page of the World Wide Dishes website}. Two call-to-action buttons were placed at the top of the page to encourage participation from site visitors. The ‘Check out our food leaderboard’ button takes site visitors to the leaderboard table, while the ‘Add your local dish’ button guides visitors through the data entry flow.}
    \label{fig:front_page}
\end{figure}

\section{Dataset collection protocol depicted through website screenshots}
\label{asec:protocol-screenshots}

\begin{figure}[H]
    \centering
    \includegraphics[width=0.7\textwidth]{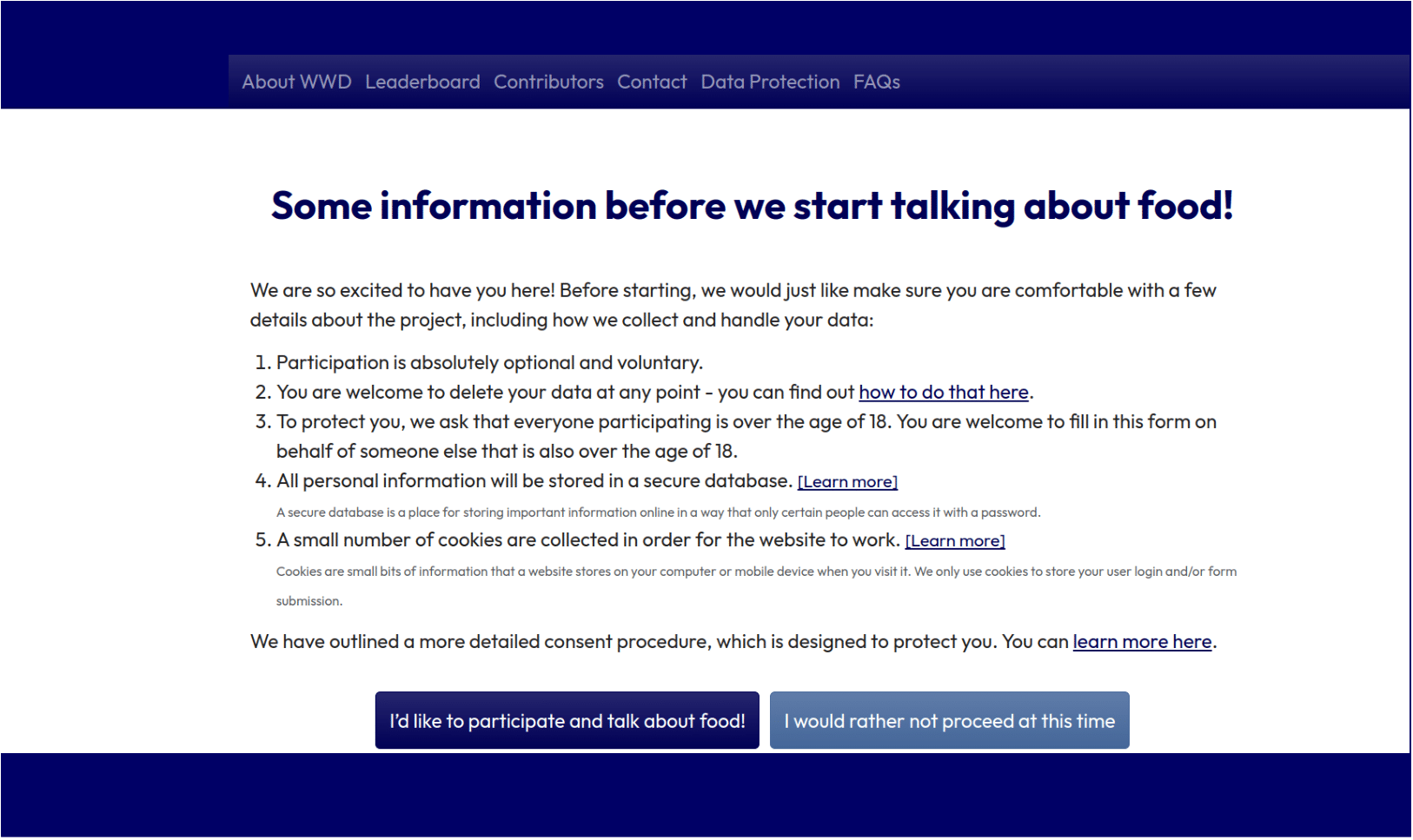}
    \caption{\textbf{Condensed informed consent page.} This page outlines important information that potential Contributors to \textbf{World Wide Dishes} must be aware of to make an informed decision about whether to contribute to the dataset. After reading through the provided list, they can choose to participate or not. The full informed consent text is shown in Figure \ref{fig:full_consent_partA} and Figure \ref{fig:full_consent_partB}. The condensed informed consent page was built after feedback from Contributors and Community Ambassadors that both the technical and essential meaning of the full form often got lost in translation to other languages.}
    \label{fig:informed_consent_new}
\end{figure}

\begin{figure}[H]
    \centering
    \includegraphics[width=0.8\textwidth]{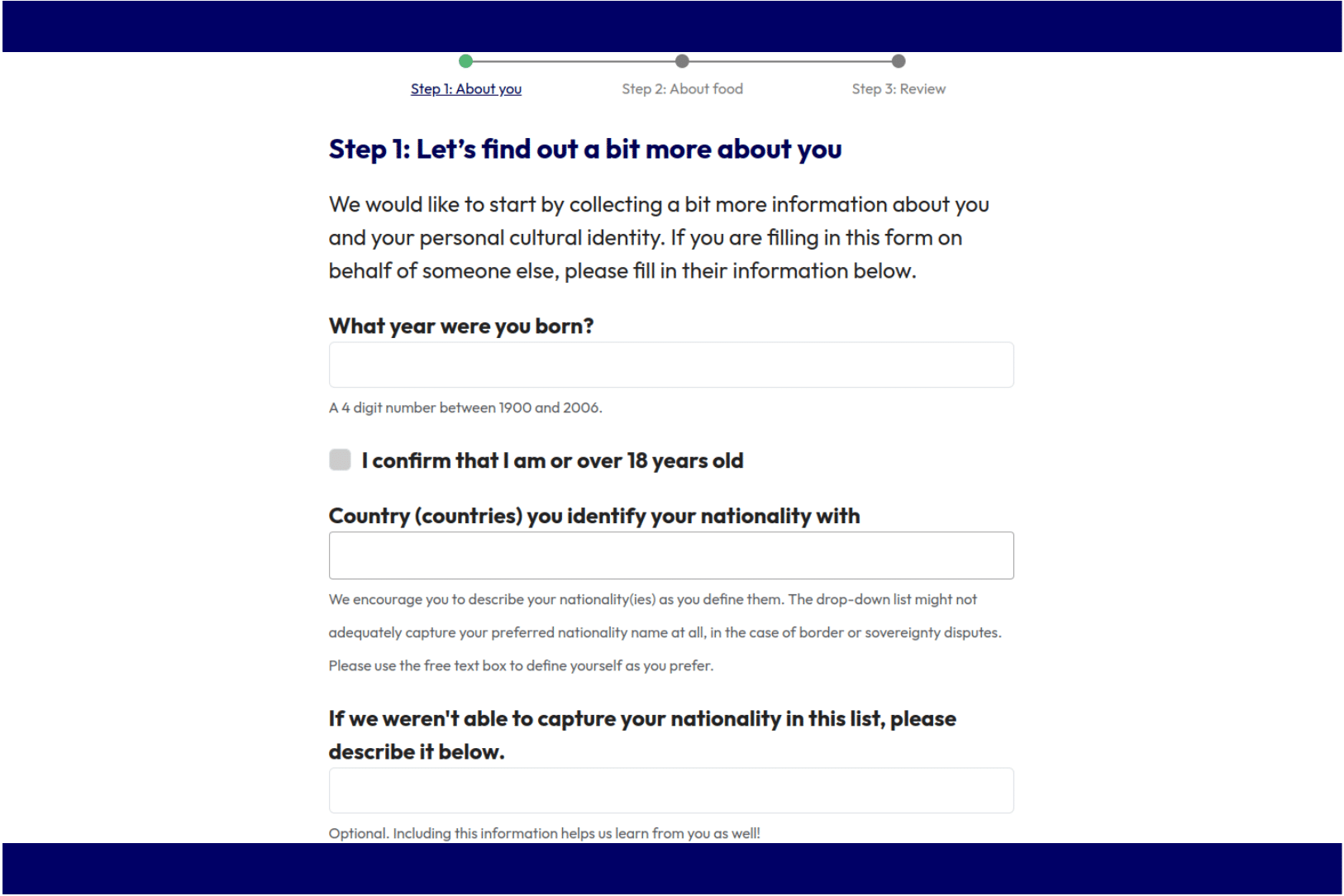}
    \caption{\textbf{``About You'' section}. After clicking on the ``I'd like to participate and talk about food!'' button on the informed consent page (Figure \ref{fig:informed_consent_new}), potential Contributors register or choose to continue as guests. Both options route the Contributors to the ``About You'' page, where they are asked to give information about their age and nationality.}
    \label{fig:personal_information}
\end{figure}

\begin{figure}[H]
    \centering
    \includegraphics[width=0.8\textwidth]{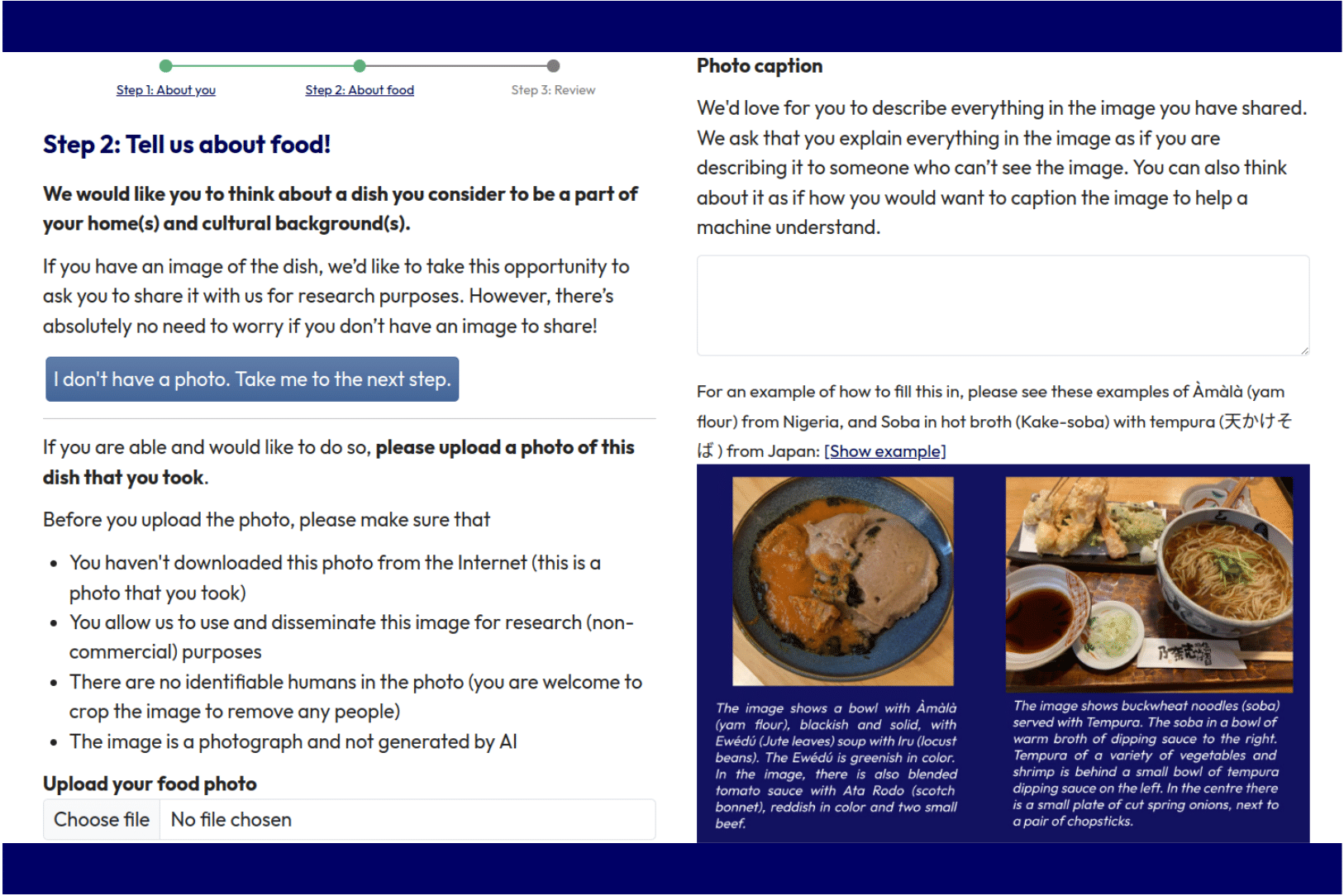}
    \caption{\textbf{``About food'' section, image upload}. In the ``About food'' section, Contributors are asked to provide data about a dish, starting with uploading an image of the dish if they have one and providing a caption. Examples of captioned images are provided to guide the data entry.}
    \label{fig:image_upload}
\end{figure}

\begin{figure}[H]
    \centering
    \includegraphics[width=0.8\textwidth]{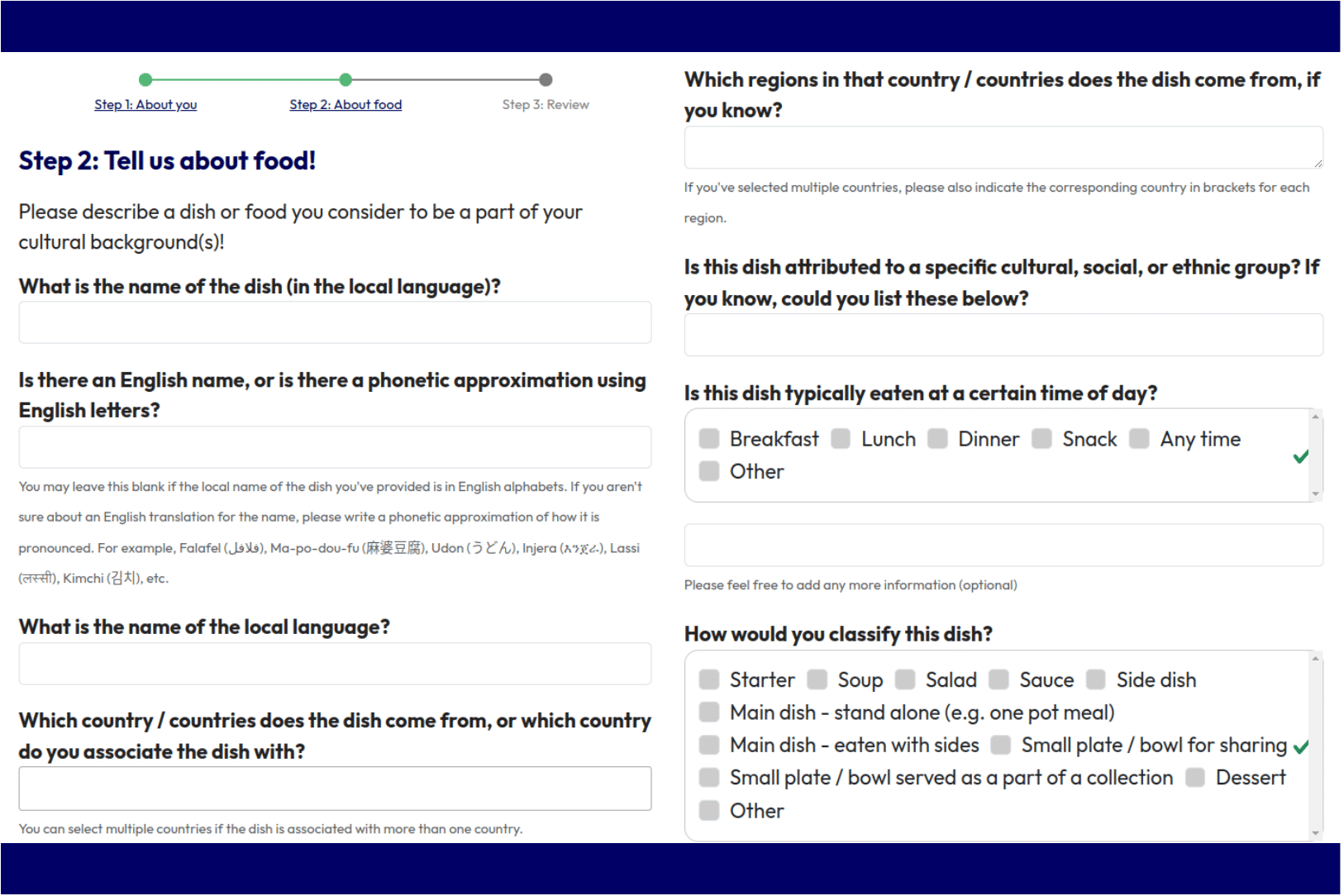}
    \caption{\textbf{About food section, continued.} Contributors are given a series of questions to answer about the dish for which they are providing information. The full list of data sought is shown in Table \ref{tab:dish_questions}.}
    \label{fig:text_upload}
\end{figure}

\begin{figure}[H]
    \centering
    \includegraphics[width=0.8\textwidth]{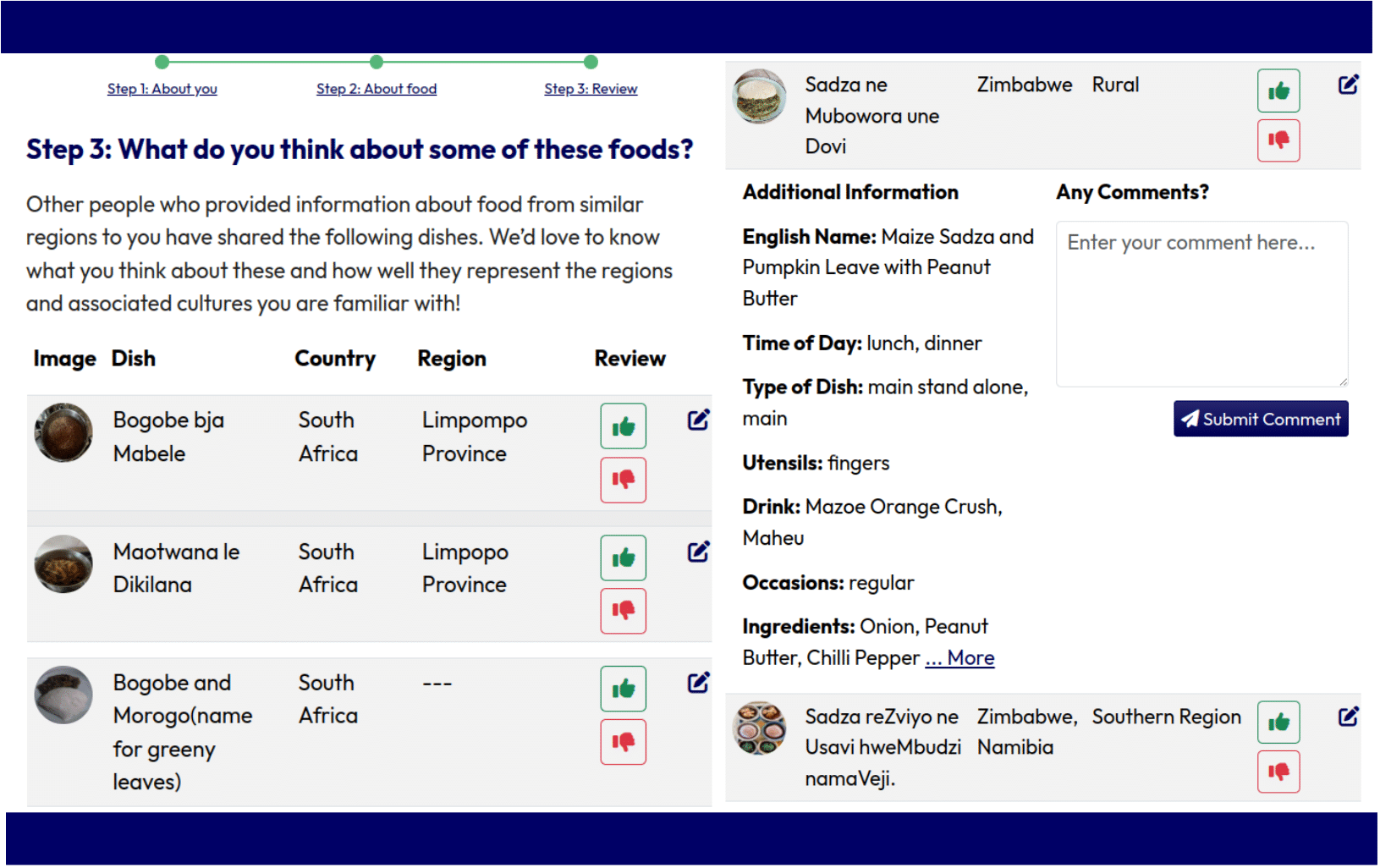}
    \caption{\textbf{``Review dishes'' section.} The final stage in the data collection process is the review step. Contributors are encouraged to review and refine data entered by others from their region.}
    \label{fig:food_review}
\end{figure}

\begin{figure}[H]
    \centering
    \includegraphics[width=0.8\textwidth]{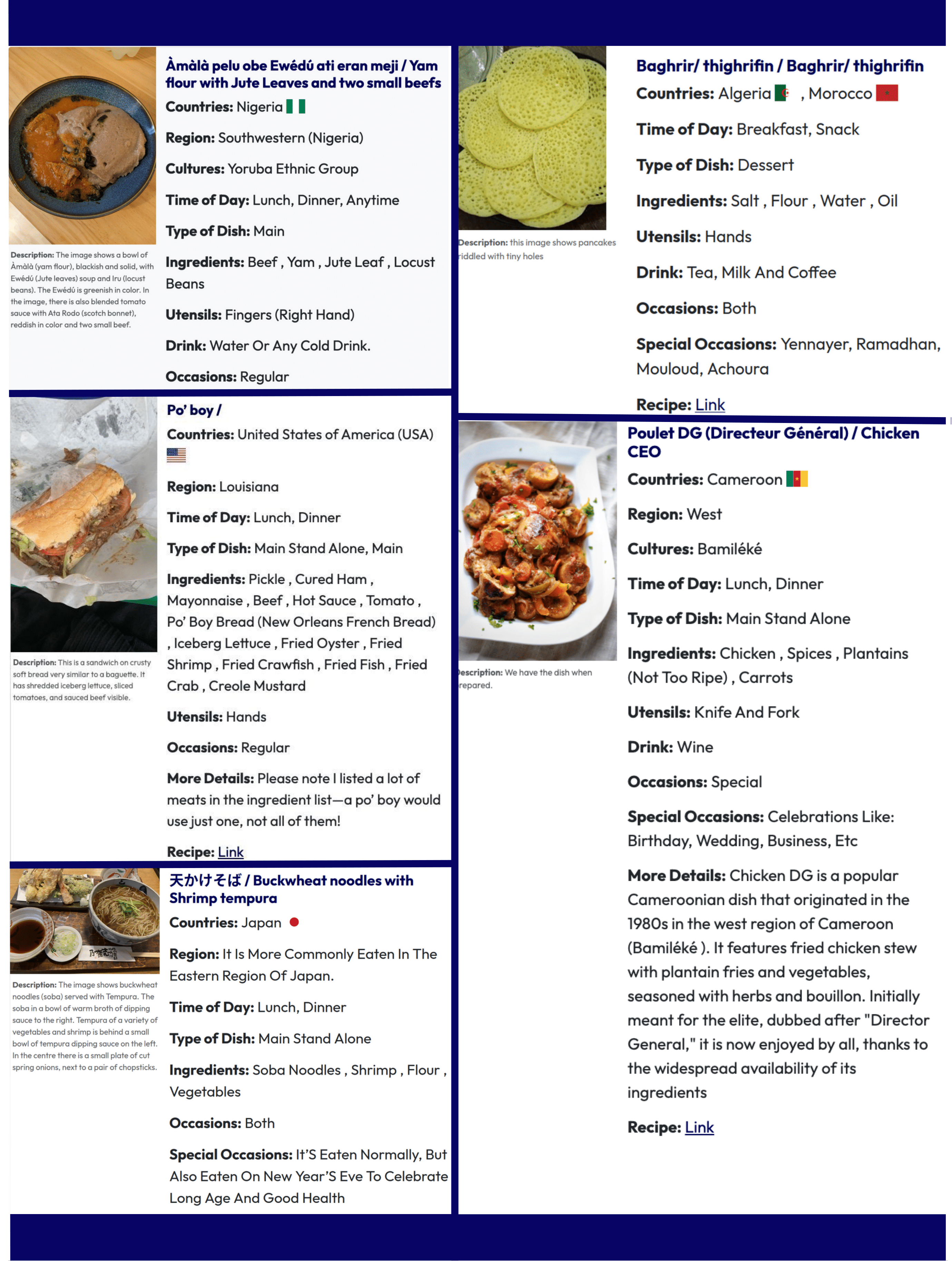}
    \caption{\textbf{A sample of entries} provided by Contributors from different regions around the world.}
    \label{fig:sample_image_dishes}
\end{figure}\label{asec:website_screenshots}

\section{The full informed consent form}\label{asec:informed_consent}
The informed consent process was designed to be informative and empowering, but still respect the different ways in which people engage with content.

\begin{figure}[H]
    \centering
    \includegraphics[width=0.8\textwidth]{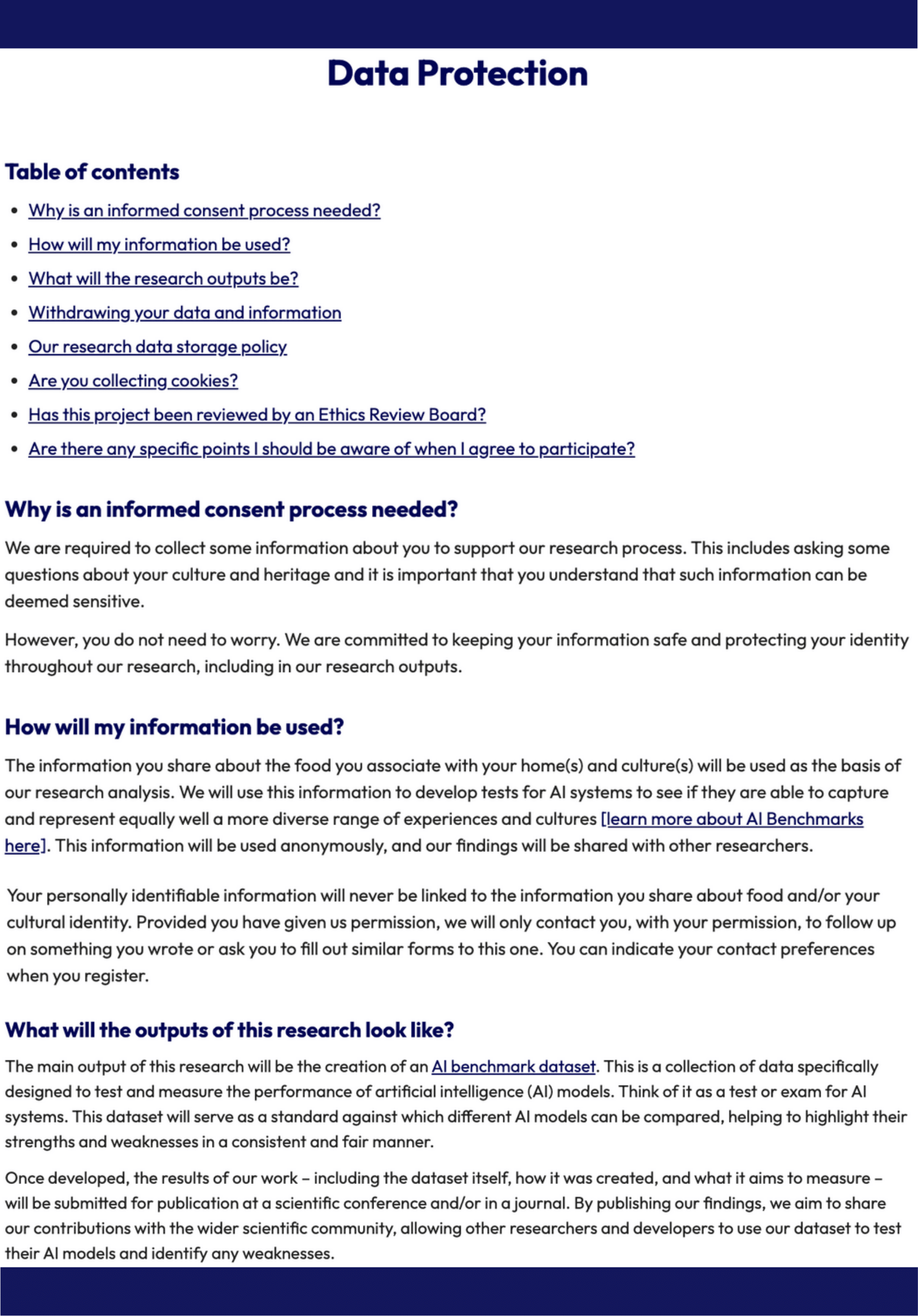}
    \caption{\textbf{Full informed consent form (Part 1 of 2).} This page provides information about why an informed consent process is needed, how Contributors' information will be used, and what research outputs will look like.}
    \label{fig:full_consent_partA}
\end{figure}

\begin{figure}[H]
    \centering
    \includegraphics[width=0.8\textwidth]{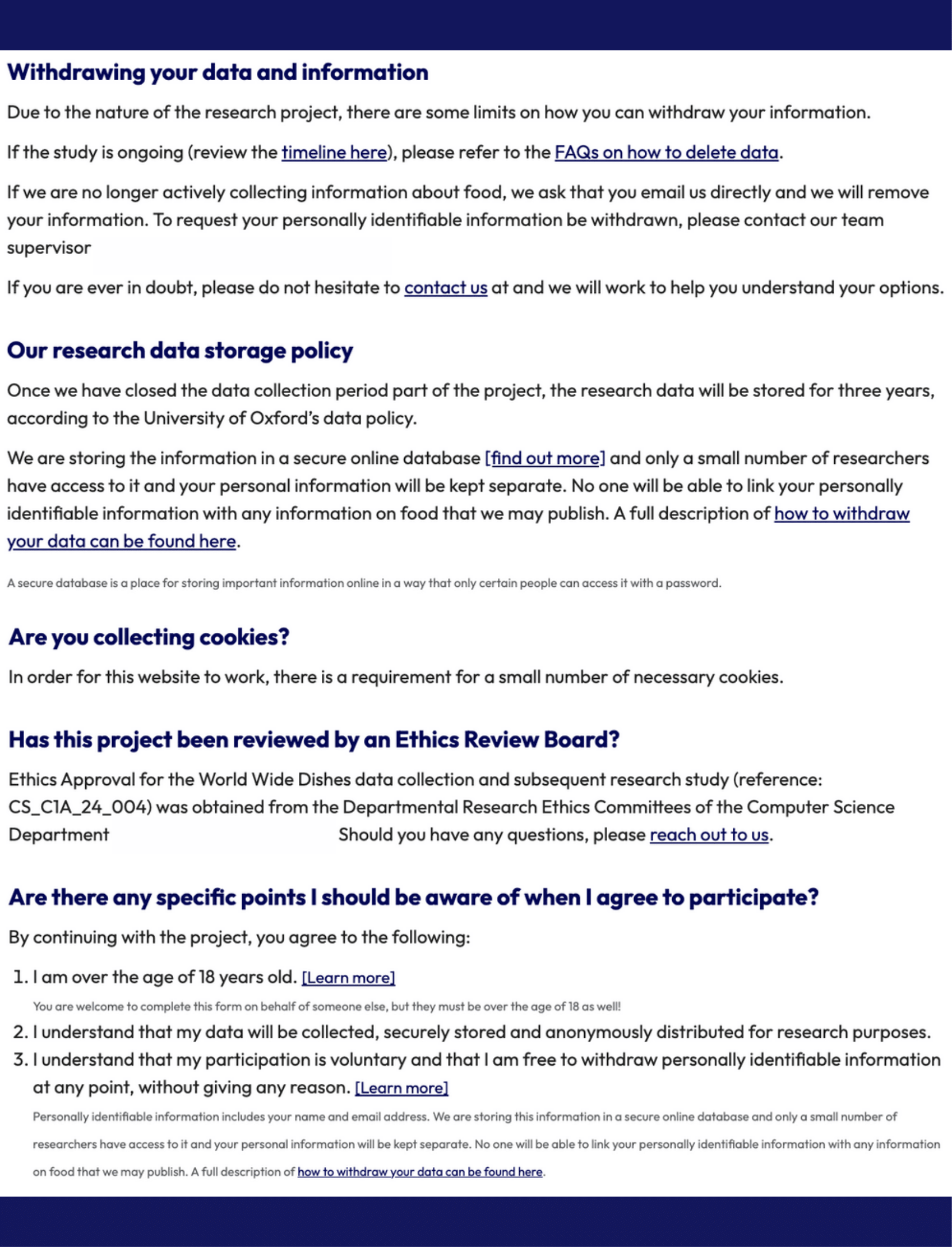}
    \caption{\textbf{Full informed consent form (Part 2 of 2).} The informed consent page further provides information about how Contributors can withdraw their data and information, the \textsc{WWD}  data storage policy, the \textsc{WWD}  ethics review board approval, and how cookies are used on the \textsc{WWD}   website.}
    \label{fig:full_consent_partB}
\end{figure}


\clearpage

\section{Metadata related to the \textsc{World Wide Dishes} dataset}\label{asec:wwd_metadata}

\subsection{Stakeholder statistics}

\begin{table}[h]
\centering
\renewcommand{\arraystretch}{1.5}

\caption{\small \textbf{Stakeholders.} We indicate the demographics for the stakeholders associated with the projects.}
\label{tab:stakeholder_tally}
\small
\begin{tabular}{p{2cm}>{\footnotesize}p{4cm}p{1.2cm}p{3cm}p{1.8cm}}
\toprule
\textbf{Stakeholder} & \textbf{Role} & \textbf{Count} & \textbf{National identities by continent} & \textbf{Age range}\\ 
\midrule
Core Organisers & The central organising team that coordinated the development and execution of World Wide Dishes. Core Organisers also acted as Community Ambassadors and Contributors & 12 & Africa, Asia, North America & - \\
Contributors and Community Ambassadors & These roles often overlapped, with Community Ambassadors supporting amplification of the WWD and Contributors submitting dishes & 162 & Africa, Asia, Europe, North America, South America, Oceania & 19-62; Mean=31.5$\pm$8.78 \\
\bottomrule
\end{tabular}
\end{table}
\vspace{-12pt}
\subsection{Continent tally}

\begin{table}[h]
\centering
\caption{\small \textbf{Number of dishes per continent.} We acknowledge that food as it has existed in history has travelled, and does not belong to specific regions exclusively.
\textsc{World Wide Dishes} has a many-to-many mapping and dishes may be tallied twice.}
\label{tab:continent_tally}
\small
\begin{tabular}{lr} 
\\
\toprule
\textbf{Continent}& \textbf{No. of Dishes}   \\ 
\midrule
Antarctica  & 0    \\
Africa  &  512     \\
Asia & 172    \\
Europe  & 58 \\
North America  &  39  \\
Oceania  &  3 \\
South America  &  11  \\

\bottomrule
\end{tabular}
\end{table}

\vspace{-12pt}
\subsection{Dish type tally}

\begin{table}[h]
\centering
\caption{\small \textbf{Number of dishes per type of meal.} We asked Contributors to associate a dish with a specific type of meal. We accept that dishes can exist in different forms, so a one-to-many mapping is used. As such, dishes can be counted more than once.}
\label{tab:type_meal_tally}
\small
\begin{tabular}{lr} 
\\
\toprule
\textbf{Type of meal}& \textbf{No. of Dishes}   \\ 
\midrule
Starter  & 80   \\
Soup & 50     \\
Salad & 12   \\
Sauce  &  22  \\
Side Dish  & 178  \\
Main dish: stand alone (e.g. a one-pot meal)  &  263 \\
Main dish: eaten with sides  &  272 \\
Small plate or bowl for sharing  &  56\\
Small plate or bowl served as a part of a collection  &  64\\
Dessert  &  107 \\
Other &  42 \\

\bottomrule

\end{tabular}
\end{table}

\section{Prompts used for post-mortem reflections}\label{asec:prompts}
The second author provided a series of prompts to the Core Organisers, who reflected upon the development and deployment of \textsc{WWD}. The second author sent these prompts in a series of text messages in a WhatsApp group chat with the COs.
\begin{itemize}
    \item How did you (as community ambassadors) interact with contributors?
    \item How did you (as community ambassadors) administer communication channels with contributors?
    \item How did you (as community ambassadors) support community contributors in engaging with WWD? 
    \item What are some memorable anecdotes you have about your focus groups you held for community contributors?
    \item How did the focus groups support data contributions from diverse populations?
    \item What issues were raised during the focus groups?
    \item How did the focus group feedback shape the design of the WWD data collection process? 
    \item How did you recruit other community ambassadors?
    \item How successful were community ambassadors in soliciting data contributions from ``hard to reach’’ populations? Did you measure success in any way?
    \item Can you share a voicenote where you reflect on how the WWD project engaged community contributors through community ambassadors? 
    \item Can you reflect on why you think contributors engaged with you?
    \item Can you describe how you introduced the project to your community?

\end{itemize}

\section{Credits for icons used in figures} \label{asec:icon_attribution}
All icons used in Figure \ref{fig:WWD_Architecture} were downloaded from Flaticon. As per Flaticon's terms of use, we provide attribution for each icon below.
\begin{itemize}
    \item `Computer: Flaticon.com'. This figure has been designed using resources from Flaticon.com..
    \item `Globe Icon: Flaticon.com'. This figure has been designed using resources from Flaticon.com.
    \item `Database: Flaticon.com'. This figure has been designed using resources from Flaticon.com.
    \item `Person Computer: Flaticon.com'. This figure has been designed using resources from Flaticon.com.
    \item `Gears: Flaticon.com'. This figure has been designed using resources from Flaticon.com.
\end{itemize}

\end{document}